\documentclass{elsart}

\input prepictex
\input pictex
\input postpictex
\begin{document}
\begin{flushright}
SNUTP 98-101\\
KIAS-P98020 \\
hep-ph/9809298\\
September 1998
\end{flushright}
\begin{frontmatter}
\title{Bounds of the mass of $Z^{\prime}$ \\
and the neutral mixing angles \\ 
in general $SU(2)_L \times SU(2)_R \times U(1)$ models}
\author[korea,kias]{Junegone Chay\thanksref{echay}}
\thanks[echay]{\tt chay@kupt.korea.ac.kr}
\author[ctp]{Kang Young Lee\thanksref{elee}}
\thanks[elee]{\tt kylee@ctp.snu.ac.kr}
\author[korea]{Soo-hyeon Nam\thanksref{enam}}
\thanks[enam]{\tt nsh@kupt2.korea.ac.kr}
\address[korea]{Department of Physics, Korea University, Seoul 136-701, 
Korea}
\address[ctp]{Center for Theoretical Physics, Seoul National
University, Seoul 151-742, Korea}
\address[kias]{Korea Institute for Advanced Study, Seoul 130-012, Korea}
\begin{abstract}
We consider phenomenological constraints on the mass
$M_{Z^{\prime}}$ and the two  mixing angles $\theta_R$ and $\xi$ of
the neutral sector in a very general class of $SU(2)_L \times SU(2)_R
\times U(1)$ models using electroweak data. We do not make any
specific assumptions such as left-right symmetry or the Higgs
structure. The analysis of the neutral sector has the advantage that
it has relatively fewer parameters compared to the charged sector
since the Cabibbo-Kobayashi-Maskawa (CKM) matrix elements in the
right-handed sector do not enter into the analysis, hence the number
of various possibilities from a big parameter space is reduced. We
utilize theoretical considerations on the masses of the gauge
particles and the mixing angles. We combine the precision electroweak
data from LEP I and the low-energy neutral-current experimental data
to constrain the parameters introduced in the model. It turns out that
$M_{Z^{\prime}}> 400$ GeV, $-0.0028 <\xi <0.0065$ with little
constraint on $\theta_R$. In the left-right symmetric theory,
$M_{Z^{\prime}}$ should be larger than 900 GeV. With these
constraints, we compare the values for $\sigma ( e^+ e^- 
\rightarrow \mu^+ \mu^-)$, $\sigma ( e^+ e^- \rightarrow
b\overline{b})$  and $A_{FB}^{\ell}$ at LEP II with  experimental
values.  

\end{abstract}

%PACS number: 12.15.Mm, 12.60.-i, 12.60.Cn

\end{frontmatter}

\section{Introduction}
The standard model with the $SU(3)_c \times SU(2)_L \times U(1)_Y$
gauge group has been successful in describing a wide range of
high energy experimental data in collider experiments and in heavy
quark decays. Almost all the experimental data can be explained within
the standard model with surprising accuracy. However it is expected
that there may be new physical phenomena beyond the standard model at
the TeV scale. Theoretically there are implications that the standard
model is an effective theory below 250 GeV, and it may be embedded in
a larger theory. There has been enormous theoretical effort in
extending the standard model such as the minimal supersymmetric
standard model \cite{mssm}, models containing many Higgs particles
\cite{higgs}, grand unified theories, and string-inspired models
\cite{string}. There are also experimental efforts to discover new 
physics effects such as supersymmetry and neutrino
oscillation \cite{neutrino}.  

As a simple extension of the standard model, we consider the question 
whether the right-handed fermions can participate in the
charged-current and the neutral-current
interactions, and if they do, with what strength. This idea can be
realized easily if we introduce the charged-current and the
neutral-current interactions for the right-handed fermions by
extending the gauge group. The simplest case arises when we choose the
gauge group as $SU(2)_L \times SU(2)_R \times U(1)$ \cite{lr}. The
left-handed fermions transform as doublets under $SU(2)_L$ and
singlets under $SU(2)_R$, with the situation reversed for the
right-handed fermions. The $U(1)$ factor is different from the
$U(1)_Y$ in the standard model. The quantum number of $U(1)$ is
proportional to $B-L$, so the model is called $SU(2)_L
\times SU(2)_R \times U(1)_{B-L}$ in some literature. The standard
model $U(1)_Y$ is obtained as a linear combination of the third
component of $SU(2)_R$ and the $U(1)$ generator.

Since we introduce an additional $SU(2)$ gauge group, it implies that
there exist three new gauge bosons: two charged and one
neutral. Therefore there are two sets of gauge bosons: the
$W_L^{\pm}$($Z_L$) belong to $SU(2)_L$ and are identical to the
$W^{\pm}$($Z$) in the standard model, while $W_R^{\pm}$($Z_R$) of
$SU(2)_R$ are new. There have 
been many theoretical and phenomenological studies of $SU(2)_L \times
SU(2)_R \times U(1)$ models \cite{lr}, and various constraints have
been presented on the mass of the new charged gauge boson $W_R^{\pm}$
and the mixing angle $\zeta$ between the two sets of charged gauge
bosons based on experimental data \cite{langacker}.  

In this paper, we consider constraints on the parameters of the
$SU(2)_L \times SU(2)R\times U(1)$ models in a very general scheme. No
assumptions are made about whether there is a left-right symmetry such
that the Lagrangian is invariant under the interchange of the
left-handed and the right-handed fermions. The left-right symmetry
implies that the gauge couplings $g_L$ and $g_R$ of the $SU(2)_L$ and
$SU(2)_R$ subgroups are equal, as well as the Yukawa couplings in each
sector. However, we release this assumption to include the possibility
of left-right asymmetric models, in which $g_L \neq g_R$.

In analyzing the charged sector, we have to carefully consider the 
origin of CP violation, and the Cabibbo-Kobayashi-Maskawa (CKM) 
matrix elements in the right-handed sector. There are many
possibilities on the CKM matrix elements according to the assumptions 
whether CP is violated spontaneously or explicitly \cite{cp}.  
Leaving aside this complication in the charged sector, we pay
attention to the neutral sector. At tree level, there is no 
flavor-changing-neutral-current process. Therefore various
possibilities of how the CKM matrix elements arise and what their
structures are both in the left- and the right-handed sector do not
enter into the analysis of the neutral sector. Due to this fact, we
can reduce specific model dependence in the analysis of the neutral
sector.

Let us consider then which parameters are necessary in the neutral
sector. There are three neutral gauge bosons: $Z_L$  
from $SU(2)_L$, $Z_R$ from $SU(2)_R$ and $B$ from $U(1)$. These gauge
fields mix to produce physical particles, one of which is
massless. There are two heavy neutral particles: $Z$ which is 
identical to $Z$ in the standard model and a new particle
$Z^{\prime}$. We need three mixing angles $\theta_R$, $\theta_W$ and
$\xi$ which describe the mixing between the gauge eigenstates and the
mass eigenstates.  The angle $\theta_W$ is identical to the Weinberg
mixing angle in the standard model. Therefore the new additional
parameters in the neutral sector are two mixing angles $\theta_R$,
$\xi$ and the mass $M_{Z^{\prime}}$ of the $Z^{\prime}$ particle. 

We consider theoretical relations among the new parameters
introduced in the $SU(2)_L \times SU(2)_R \times U(1)$ model. The
gauge boson masses have definite relations between the charged
sector and its neutral counterpart. There are also relations between
the small mixing angle $\xi$ and the mass ratio
$M_Z^2/M_{Z^{\prime}}^2$, as there is a similar relation between the
mixing angle $\zeta$ and the ratio $M_W^2/M_{W^{\prime}}^2$ in the
charged sector. We utilize these relations along with the experimental
data to constrain the parameters in the model.

The main points in this paper are to probe the parameter space spanned
by $\theta_R$, $\xi$ and $M_{Z^{\prime}}^2$ and to look for the region
that is allowed by current experimental data. We first 
consider electroweak precision data from LEP I. Since LEP I performs
the experiment at the $Z$ peak, the corrections do not depend on
$M_{Z^{\prime}}^2$ explicitly. Therefore the LEP I data constrain only
$\theta_R$ and $\xi$. We combine these bounds with the low-energy
neutral-current data to constrain $\theta_R$, $\xi$ and
$M_{Z^{\prime}}$. With these results, we predict the cross
sections for $e^+ e^- \rightarrow \mu^+ \mu^-$, $e^+ e^- \rightarrow
b\overline{b}$ and the leptonic forward-backward asymmetry
$A_{FB}^{\ell}$ at LEP II energies and compare with experimental data.

This paper is organized as follows: In Section 2 we formulate the
$SU(2)_L \times SU(2)_R \times U(1)$ model. After reviewing the
charged sector briefly, we describe in detail the neutral gauge bosons
and the interactions with them. We introduce three mixing angles
explicitly to diagonalize the neutral gauge boson mass matrix. In
obtaining the current interactions, we note the fact that
$M_{Z^{\prime}}\gg M_Z$ and the mixing angle $\xi$, which describes
the mixing of the two massive gauge bosons, is expected to be small
($\xi \sim M_Z^2/M_{Z^{\prime}}^2$). We express all the quantities to
first order in $\xi$ and $M_Z^2/M_{Z^{\prime}}^2$ in order for the 
numerical estimates to be consistent. Using the information on the
mixing, we obtain the current interactions. In Section 3 we consider
theoretical relations between $M_{Z^{\prime}}^2$, $M_{W^{\prime}}^2$,
and $M_Z^2$, $M_W^2$. We get the bounds for the mixing angles $\zeta$
and $\xi$ in terms of the mass ratios. 

In Section 4, we use experimental data to constrain the parameters in
the theory. There are many physical quantities observed at LEP I. The
useful quantities in our analysis are the leptonic
decay width $\Gamma (Z\rightarrow \ell^+ \ell^-)$ and the
leptonic forward-backward asymmetry $A_{FB}^l$. We combine the
constraints obtained from LEP I with low-energy neutral-current
data. For low-energy neutral-current interactions, we consider
neutrino-electron scattering, neutrino-hadron scattering, polarized
electron-hadron scattering and the atomic parity violation.
With the bounds obtained from the combined data of LEP I and the
low-energy data, we consider the cross sections $\sigma (e^+ e^-
\rightarrow \mu^+ \mu^-)$, $\sigma (e^+ e^- \rightarrow
b\overline{b})$ and the leptonic forward-backward asymmetry $A_{FB}^l$
at LEP II  energies. In Section 5, a summary of the analysis is given
and a conclusion is presented. 

\section{$SU(2)_L \times SU(2)_R \times U(1)$ model}
\subsection{General structure}
Consider the theory based on the electroweak gauge group $SU(2)_L
\times SU(2)_R \times U(1)$. Such theory has been extensively
investigated both as a simple generalization of the $SU(2)_L \times
U(1)_Y$ model and as possible intermediate stages in grand unified
schemes such as $SO(10)$. In constructing the $SU(2)_L \times SU(2)_R
\times U(1)$ theory, it is appealing to impose a discrete
left-right symmetry so that parity is restored at a higher energy
scale above the weak scale. This additional symmetry simplifies the
structure of the Lagrangian. However, it is not required by the
structure of the extended gauge group. Furthermore left-right
symmetric theories encounter difficulties in the context of grand
unified models or cosmology \cite{lrsym}.  We do not impose the
left-right symmetry from the outset so that our model includes a class
of asymmetric left-right models, and we will work within the  
general framework of the $SU(2)_L \times SU(2)_R \times U(1)$.

We start with the extended gauge group  $SU(2)_L \times SU(2)_R \times
U(1)$ which breaks down to $SU(2)_L \times U(1)_Y$ at the energy scale
$v_R$, much larger than the weak scale. The remaining gauge group
is identified as that of the standard model and it finally cascades
down to $U(1)_{\mathrm{em}}$. The covariant derivative is defined as 
\begin{equation}
D_{\mu} = \partial_{\mu} + ig_L W_{L\mu}^a T_{La} +ig_R W_{R\mu}^a
T_{Ra} + ig_1 B_{\mu}S,
\end{equation}
where $T_{L,R}^a$ are $SU(2)_{L,R}$ generators and $S$ is the $U(1)$
generator. $g_L$, $g_R$ and $g_1$ are the gauge coupling constants
for the corresponding gauge groups. The representations of quarks and
leptons under the gauge group $SU(2)_L \times SU(2)_R \times U(1)$ are
given as
\begin{eqnarray}
q_L^{\prime} &=& \left( \begin{array}{c}
                     u^{\prime} \\ d^{\prime}
                       	\end{array}
                  \right)_L \sim (2,1)^{1/6}, \ 
q_R^{\prime} = \left( \begin{array}{c}
                     u^{\prime} \\ d^{\prime}
                       	\end{array}
                  \right)_R \sim (1,2)^{1/6}, \nonumber \\
l_L^{\prime} &=& \left( \begin{array}{c}
                     \nu^{\prime} \\ e^{\prime}
                       	\end{array}
                  \right)_L \sim (2,1)^{-1/2}, \
l_R^{\prime} = \left( \begin{array}{c}
                     \nu^{\prime} \\ e^{\prime}
                       	\end{array}
                  \right)_R \sim (1,2)^{-1/2}.
\label{rep}
\end{eqnarray}
Under the strong gauge group $SU(3)_c$, quarks transform as triplets
while leptons transform as singlets. The primed fields in
Eq.~(\ref{rep}) denote gauge eigenstates rather than mass eigenstates. 

In order to invoke spontaneous symmetry breaking, we introduce the
scalar field
\begin{equation}
\Phi = \left( \begin{array}{cc}
             \phi_1^0 & \phi_1^+ \\
             \phi_2^- & \phi_2^0
              \end{array}
       \right) \sim  (2,\overline{2})^0,
\end{equation}
which acquires the vacuum expectation value (VEV) 
\begin{equation}
\langle \Phi \rangle = \left( \begin{array}{cc}
             k & 0 \\
             0 & k^{\prime}
              \end{array}
       \right).
\end{equation}
In general $k$, $k^{\prime}$ can be complex. This VEV generates
fermion masses in the Yukawa sector after the spontaneous symmetry
breaking.   

We need to include additional scalar fields into our theory to
implement the symmetry breaking pattern $SU(2)_L \times SU(2)_R \times
U(1)_S \rightarrow SU(2)_L \times U(1)_Y \rightarrow
U(1)_{\mathrm{em}}$. There are a number of choices for how these
scalars transform under the full symmetry group. However we choose the
simplest case which involves two doublet fields,
\begin{equation}
\chi_L = \left( \begin{array}{c} 
               \chi_L^+ \\
               \chi_L^0
	       \end{array}
       \right) \sim (2,1)^{1/2}, \ 
\chi_R = \left( \begin{array}{c} 
               \chi_R^+ \\
               \chi_R^0
	       \end{array}
       \right) \sim (1,2)^{1/2}
\end{equation}
which acquire the real VEVs
\begin{equation}
\langle \chi_L\rangle = \left( \begin{array}{c} 
               0 \\
               v_L
	       \end{array}
       \right),\  
\langle\chi_R \rangle = \left( \begin{array}{c} 
               0 \\
               v_R
	       \end{array}
       \right).
\end{equation}
Though $\chi_L$ is not necessary for the desired structure of the
symmetry breaking, we introduce it anyway along with $\chi_R$ so that
our model can accommodate left-right symmetric models. 

The Lagrangian for the scalar fields is given by
\begin{eqnarray}
L_{\mathrm{scalar}} &=& \mbox{tr}\Bigl( (D^{\mu} \Phi)^{\dagger}
D_{\mu} \Phi \Bigr) + (D^{\mu} \chi_L)^{\dagger} D_{\mu} \chi_L +
(D^{\mu} \chi_R)^{\dagger} D_{\mu} \chi_R \nonumber \\
&&-V(\Phi,\chi_L,\chi_R),
\label{scalar}
\end{eqnarray}
where the covariant derivatives acting on the scalar fields are
defined as
\begin{eqnarray}
&&D^{\mu} \Phi = \partial^{\mu} \Phi + ig_L W^a_{L\mu} T_{La} \Phi - 
ig_R W^a_{R\mu} \Phi T_{Ra}, \nonumber \\
&&D^{\mu} \chi_{L,R} = \partial^{\mu} \chi_{L,R} + ig_{L,R}
W^a_{L,R\mu} T_{L,Ra} \chi_{L,R}.
\end{eqnarray}
The scalar potential $V(\Phi,\chi_L, \chi_R)$ is constructed such that
the spontaneous symmetry breaking occurs and the minimum of the
potential produces the desired VEVs \cite{senja}. We will not go
further into the detail of how this is possible. After the spontaneous
symmetry breakdown, the kinetic energy in Eq.~(\ref{scalar}) induces
gauge boson masses. 

We can also have a variation of the contents of the scalar fields. For 
example, we can introduce a scalar triplet in order to produce
Majorana neutrino mass. In addition to producing Majorana neutrino
masses, the scalar triplet can affect the gauge boson mass
matrix. However, the contribution of the triplet for the gauge boson
masses amounts to effectively changing the VEVs of the existing scalar
fields, hence not affecting the analysis of the gauge sector. 
   
\subsection{Charged gauge bosons}
After the spontaneous symmetry breakdown, the scalar Lagrangian in
Eq.~(\ref{scalar}) generates gauge-boson masses. The mass-squared
matrix for the charged gauge bosons can be written as
\begin{equation}
M_{W^{\pm}}^2 = \left( \begin{array}{cc}
                g_L^2 (v_L^2 + K^2)/2 & -g_L g_R k^*
                k^{\prime} \\
               -g_L g_R kk^{\prime *} & g_R^2(v_R^2 + K^2)/2
		\end{array}
              \right) \equiv 
              \left( \begin{array}{cc}
              M_L^2& M_{LR}^2 e^{i\alpha} \\
              M_{LR}^2 e^{-i\alpha}& M_R^2 
		     \end{array}
             \right),
\end{equation}
where $K^2 = |k|^2 + |k^{\prime}|^2$ and $\alpha$ is the phase of $k^*
k^{\prime}$. 

We introduce the mixing angle $\zeta$ between the two gauge
eigenstates defined by
\begin{equation}
\tan 2\zeta = -\frac{2M_{LR}^2}{M_R^2 - M_L^2}.
\label{zeta}
\end{equation}
Then the eigenvalues can be written in terms of $\zeta$ as
\begin{eqnarray} 
M_W^2 &=& M_L^2 \cos^2 \zeta +M_R^2 \sin^2 \zeta +M_{LR}^2 \sin
2\zeta, \nonumber \\
M_W^{\prime 2} &=& M_L^2 \sin^2 \zeta + M_R^2 \cos^2 \zeta -M_{LR}^2
\sin 2\zeta,
\label{wmass}
\end{eqnarray}
and the corresponding eigenvectors are
\begin{equation}
\left( \begin{array}{c}
W^+ \\
W^{\prime +} 
       \end{array}
\right) = \left( \begin{array}{cc}
        \cos \zeta & e^{-i\alpha} \sin \zeta \\
        -\sin \zeta & e^{-i\alpha} \cos \zeta
		 \end{array}
        \right)
\left( \begin{array}{c}
     W_L^+ \\
      W_R^+
       \end{array}
\right).
\end{equation}
The $W^+$ and $W^{\prime +}$ fields are the physical charged gauge
bosons in the $SU(2)_L \times SU(2)_R \times U(1)$ theory. For $v_R
\gg v_L, |k|, |k^{\prime}|$, the mass $M_W^{\prime}$ ($M_W$) is almost
the mass of the $W_R$ ($W_L$) particle since the $W_L$-$W_R$ mixing
angle $\zeta$ is small. 

The two parameters $M_W^{\prime}$ and $\zeta$ are those appearing in
the charged sector and they are restricted by a number of low-energy
phenomenological constraints along with the CKM matrix elements in the
right-handed sector. Conservative numerical estimates for the bounds
lie in the range \cite{langacker}  
\begin{equation}
M_W^{\prime} > 300 \ \mbox{GeV}, \ \ |\zeta| < 0.075.
\label{charge}
\end{equation}
These bounds can differ depending on the form of the CKM matrix in the
right-handed sector and the type of the right-handed neutrino. (For
reference, see Tables I and II in Ref.~\cite{langacker}.) The quoted
bounds in Eq.~(\ref{charge}) are the weakest bounds and, in other
specific cases, the bounds get tighter. For example, for the
left-right symmetric case, the bound for the $Z^{\prime}$ mass can go
up as high as 1.4 TeV. These constraints have been obtained from the 
$K_L$-$K_S$ mass difference, $B_d \overline{B}_d$ mixing, semileptonic
$b$ decays and the neutrinoless double beta decay. 
 
\subsection{Neutral gauge bosons}
The Lagrangian in Eq.~(\ref{scalar}) also produces the masses of the
neutral gauge bosons after the spontaneous symmetry breakdown. In a
similar way, we can construct the mass-squared matrix for the neutral
gauge bosons. The diagonalization of the mass matrix in the neutral
sector is more complicated since there are three neutral gauge
particles $W_{L3}$, $W_{R3}$ and $B$ and the mass matrix becomes a
$3\times 3$ matrix. The mass-squared matrix for the
neutral gauge bosons is given by 
\begin{equation} 
M^2 = \left( \begin{array}{ccc}
g_L^2(v_L^2 + K^2)/2 & - g_L g_R K^2/2 & -g_L g_1 v_L^2/2 \\
-g_L g_R K^2/2 & g_R^2 (v_R^2 + K^2)/2 & -g_R g_1 v_R^2/2 \\
-g_L g_1 v_L^2/2 & -g_R g_1 v_R^2/2 & g_1^2
(v_L^2 + v_R^2)/2 
	     \end{array}
\right).
\label{neumass}
\end{equation}

The mass-squared matrix in Eq.~(\ref{neumass}) is a real symmetric
$3\times 3$ matrix. In order to diagonalize it, we need a real
orthogonal matrix which can be parameterized in terms of three 
Euler-type angles. We introduce $\theta_R$, $\theta_W$ and $\xi$ as
three mixing angles in diagonalizing the mass matrix. The derivation
of obtaining the mass eigenstates in terms of the gauge eigenstates
and the introduction of three mixing angles are presented in Appendix
in detail.   

The mass eigenstates can be expressed in terms of the gauge
eigenstates as   
\begin{eqnarray} 
A^{\mu} &=& \sin \theta_W W_{L3}^{\mu} + \cos \theta_W \sin \theta_R
W_{R3}^{\mu} +\cos \theta_W \cos \theta_R B^{\mu}, \nonumber \\
Z^{\mu} &=& \cos \xi \cos \theta_W W_{L3}^{\mu} +(\sin \xi \cos
\theta_R -\cos \xi \sin \theta_W \sin \theta_R) W_{R3}^{\mu} \nonumber
\\
&&- (\cos \xi \sin \theta_W \cos \theta_R + \sin \xi \sin \theta_R)
B^{\mu} ,\nonumber \\
Z^{\prime \mu} &=& -\sin \xi \cos \theta_W W_{L3}^{\mu} + (\cos \xi
\cos \theta_R + \sin \xi \sin \theta_W \sin \theta_R) W_{R3}^{\mu}
\nonumber \\
&&+ (\sin \xi \sin \theta_W \cos \theta_R - \cos \xi \sin \theta_R)
B^{\mu}.
\label{mixings}
\end{eqnarray}
The  mixing angles $\theta_R$ and $\theta_W$ are defined as 
\begin{equation}
\sin \theta_R = \frac{g_1}{\sqrt{g_1^2 + g_R^2}}, \
\sin \theta_W = \frac{g_1 g_R}{\sqrt{g_1^2 g_R^2 + g_L^2 g_1^2 +
g_L^2 g_R^2}}.
\label{mangle}
\end{equation}
As noted in Appendix, the mixing angle $\theta_W$ between the $A$
and the $Z$ fields corresponds to the Weinberg mixing angle in the
standard model. The mixing angle $\xi$ is defined in Appendix and it
is very small.

The eigenvalues for the mass eigenstates are given by
\begin{eqnarray}
M_A^2 &=& 0, \nonumber \\
M_Z^2 &\approx& \frac{1}{2} \frac{g_L^2}{c_W^2} (K^2 + v_L^2)
- \frac{g_L^2}{2c_W^2} \frac{( c_R^2 K^2 - s_R^2 v_L^2)^2}{v_R^2},
\nonumber \\     
M_{Z^{\prime}}^2 &\approx& \frac{1}{2} (g_1^2 + g_R^2 ) v_R^2 +
\frac{1}{2} (g_R^2 c_R^2 K^2 + g_1^2 s_R^2 v_L^2),
\end{eqnarray}
where $c_i = \cos i$, $s_i = \sin i$ with $i=\theta_W$, $\theta_R$. 
The massless field $A$ corresponds to the photon field. $Z^{\prime}$
is the heaviest since its mass is proportional to $v_R$ ($v_R \gg v_L,
k, k^{\prime}$). Note that, at order $O(1/v_R^2)$,  $M_Z$ decreases
while $M_{Z^{\prime}}$ increases compared to the leading values. This
is a general feature of quantum mechanics that, if there is mixing in
a two-level system, the mixing widens the energy gap between the
unperturbed states.

In summary the neutral gauge sector is described by three mixing
angles, one of which is the Weinberg mixing angle. And there is the
mass $M_{Z^{\prime}}$ of the new heavy particle. These parameters
$\theta_R$, $\xi$ and $M_{Z^{\prime}}$ are going to be restrained from 
experimental data. 

\subsection{Current interactions}

Here we write down the current interactions in the $SU(2)_L \times
SU(2)_R \times U(1)$ model. Though we will not consider the
charged-current interactions, we list them for completeness.  
Now that we have physical, mass eigenstates for the gauge bosons, we
can write down the interaction of fermions with gauge particles. First
we write down the fermion fields in terms of mass eigenstates. After
the symmetry breakdown, fermions get masses and the mass eigenstates
are related to the gauge eigenstates by the following unitary
transformations
\begin{equation}
u_L^{\prime} = S_u u_L, \ u_R^{\prime}= T_u u_R, \ d_L^{\prime} = S_d
d_L, \ d_R^{\prime} = T_d d_R, 
\end{equation}
where the primed fields are gauge eigenstates and the unprimed fields
are mass eigenstates. Let $V_L = S_u^{\dagger} S_d$ and $V_R =
T_u^{\dagger} T_d$. $V_L$ is the usual CKM matrix in the standard
model and $V_R$ is its counterpart in the right-handed sector.  

The charged-current interaction is given by
\begin{eqnarray} 
L_{\mathrm{CC}} &=& \frac{1}{\sqrt{2}} (\overline{u},
\overline{c}, \overline{t}) \Bigl[ ~ \FMSlash{W}_{+} \Bigl( -g_L \cos 
\zeta V_L P_- - g_R \sin \zeta e^{i\alpha} V_R P_+ \Bigr) \nonumber \\ 
&&+~ \FMSlash{W}_{+}^{\prime} \Bigl( g_L \sin \zeta V_L P_- -g_R \cos 
\zeta e^{i\alpha} V_R P_+ \Bigr) \Bigr] 
\left(\begin{array}{c}
d\\
s\\
b
      \end{array}
\right) + \mathrm{h.c.},
\end{eqnarray}
where $P_{\mp} = (1\mp \gamma_5)/2$ are the left- and right-handed
projection operators respectively. The leptonic charged-current
interactions can be written in a similar way. Since the analysis on
$\zeta$ and $M_{W^{\prime}}$ in the charged sector has been
extensively investigated in Ref.~\cite{langacker}, we will not
consider it here any more. 

In the neutral-current interaction, there are no flavor-changing
neutral currents at tree level. Therefore the CKM matrices $V_L$ and
$V_R$ do not appear and the Glashow-Illiopoulos-Maiani mechanism still
works in the $SU(2)_L \times SU(2)_R \times U(1)$ model. The
neutral-current interaction is written as
\begin{eqnarray}
L_{\mathrm{NC}} &=& -e \overline{\psi} \FMSlash{A} \Bigl[ (T_{L3} +S)
P_- + (T_{R3} +S) P_+ \Bigr] \psi \nonumber \\
&&- \overline{\psi} \FMSlash{Z}\Bigl[ \Bigl( g_L c_W c_{\xi} T_{L3} -
g_1 (c_R s_W c_{\xi} + s_R s_{\xi} ) S\Bigr) P_- \nonumber \\
&&\ \ + \Bigl( g_R (c_R s_{\xi} - s_R s_W c_{\xi}) T_{R3} - g_1 (c_R
s_W c_{\xi} + s_R s_{\xi} )S \Bigr) P_+ \Bigr] \psi \nonumber \\
&&- \overline{\psi} \FMSlash{Z}^{\prime} \Bigl[ \Bigl(-g_L c_W
s_{\xi} T_{L3} + g_1 (c_R s_W s_{\xi} - s_R c_{\xi})S \Bigr) P_-
\nonumber \\
&&\ \ + \Bigl( g_R (c_R c_{\xi} + s_R s_W s_{\xi} ) T_{R3} + g_1 (c_R
s_W  s_{\xi} - s_R c_{\xi})S \Bigr) P_+ \Bigr] \psi,
\label{neutcur}
\end{eqnarray}
Here $\psi$ denotes fermion fields and $c_{\xi} = \cos \xi$ and 
$s_{\xi} = \sin \xi$.

From Eq.~(\ref{neutcur}), the electric charge operator $Q$ can be
obtained by looking at the coupling of fermions with the photon. It is
given by  
\begin{equation} 
Q = T_{L3} + T_{R3} +S.
\end{equation}
And we also introduce the electromagnetic coupling constant $e$, and  
the relations among the coupling constants are given by
\begin{equation}
e=g_L s_W = g_R s_R c_W = g_1 c_R c_W = \frac{g_1 g_R g_L}{\sqrt{g_1^2 
g_R^2 + g_L^2 g_1^2 + g_L^2 g_R^2}}.
\label{echarge}
\end{equation}

The current coupled to the $Z$ field to first order in $\xi$ is given
by 
\begin{equation}
J_Z^{\mu} = \frac{e}{s_W c_W} \overline{\psi} \gamma^{\mu} \Bigl
[ T_{L3} - s_W^2 Q + \xi s_W \Bigl( t_R (T_{L3} -Q) + (t_R +
\frac{1}{t_R}) T_{R3} \Bigr)\Bigr]\psi,  
\end{equation}
where $t_R = s_R/c_R$. And the current coupled to $Z^{\prime}$ to
zeroth order in $\xi$ is 
\begin{equation}
J_{Z^{\prime}}^{\mu} = \frac{e}{c_W} \overline{\psi} \gamma^{\mu}
\Bigl( t_R (T_{L3} -Q) + (t_R + \frac{1}{t_R}) T_{R3} \Bigr) \psi. 
\end{equation}
Since those processes mediated by $J_{Z^{\prime}}^{\mu}$ will be
already suppressed by $1/M_{Z^{\prime}}^2$, we do not include the term
proportional to $\xi$.

\section{Theoretical considerations}

Before we use experimental data to constrain the parameters in our
model, it is useful to consider the structure of the theory. There are
a few interesting relations between the masses of the gauge bosons. 
When the mixing angles $\zeta$ and $\xi$ are small, we
can estimate these mixing angles in terms of the mass ratios of the
gauge bosons. Though the bounds obtained from theoretical
considerations may not be helpful in obtaining strong constraints on
the parameters, it is worthwhile to notice the interwoven structure of 
the theory and it will give rough estimates of the parameters.

\subsection{Masses of $Z^{\prime}$ and $W^{\prime}$}
The exact masses of $W^{\prime}$ and $Z^{\prime}$ are expressed in
terms of mixing angles with VEVs as in Eqs.~(\ref{wmass}),
(\ref{zmass}). Since we assume that $v_R \gg k, k^{\prime}, v_L$,
we can express the masses in a power series with respect to $1/v_R^2$
(or equivalently in powers of $\xi$). The approximate masses are given
by
\begin{equation}
M_{W^{\prime}}^2 \approx \frac{1}{2} g_R^2 v_R^2 \Bigl( 1+
\frac{K^2}{v_R^2} \Bigr), \ \ 
M_{Z^{\prime}}^2 \approx \frac{1}{2} (g_1^2 + g_R^2) v_R^2 \Bigl( 1+
\frac{c_R^4 K^2 + s_R^4 v_L^2}{v_R^2}\Bigr). 
\end{equation}

The ratio of these masses is given by
\begin{equation}
\frac{M_{Z^{\prime}}^2}{M_{W^{\prime}}^2} \approx \frac{1}{c_R^2}
\Bigl[ 1+ \frac{s_R^2}{v_R^2} \Bigl(s_R^2 v_L^2 - (1+c_R^2) K^2 \Bigr)
\Bigr],
\end{equation}
which leads to the relation
\begin{equation} 
M_{W^{\prime}}^2 = M_{Z^{\prime}}^2 c_R^2 +O(K^2/v_R^2).
\label{cust2}
\end{equation}
From Eq.~(\ref{cust2}), we conclude that $M_{Z^{\prime}}$ is larger
than $M_{W^{\prime}}$.

This is the relation between $M_{W^{\prime}}^2$ and
$M_{Z^{\prime}}^2$. It is similar to the relation 
between $M_W^2$ and $M_Z^2$ ($M_W^2 = M_Z^2 c_W^2$) in the standard
model. If we define the $\rho$ and $\rho^{\prime}$ parameters in the
$SU(2)_L \times SU(2)_R \times U(1)$ model as $\rho \equiv M_W^2/M_Z
c_W^2$ and $\rho^{\prime} \equiv M_{W^{\prime}}^2 /M_{Z^{\prime}}^2
c_R^2$ at tree level, they are close to 1 up to order
$O(K^2/v_R^2)$. That is,  
\begin{equation}
\rho \approx \rho^{\prime} = 1 +O(K^2/v_R^2).
\end{equation}
In the standard model, since the scalar field has a larger symmetry
than the original $SU(2)_L$ symmetry, this
extra symmetry requires $\rho =1$. This is called the custodial
symmetry. In the $SU(2)_L \times SU(2)_R \times U(1)$ model, 
there are two kinds of custodial symmetries when we construct the
scalar fields as we have done. The relation $\rho^{\prime} \approx 1$
is the result of the  additional custodial symmetry. The correction to
$\rho = \rho^{\prime} =1$ is due to the mixing $\xi$ of order
$O(K^2/v_R^2)$ between the two massive neutral states.

From the relation $M_{W^{\prime}} \approx M_{Z^{\prime}} c_R$, we only
know that $M_{Z^{\prime}} > M_{W^{\prime}}$. In the special
case with the left-right symmetry ($g_L =g_R$), we can have a stronger
bound. Using the relation in Eq.~(\ref{echarge}), when $g_L = g_R$,
$\sin \theta_R = \tan \theta_W$. Using the value $\sin^2 \theta_W =
0.231$, we get $\theta_W = 28.7^{\circ}$ and
$\theta_R=33.2^{\circ}$, resulting in $\cos \theta_R < \cos \theta_W$. 
Therefore we get the bound
\begin{equation}
M_{Z^{\prime}} = \frac{M_{W^{\prime}}}{\cos \theta_R}  >
\frac{M_{W^{\prime}}}{\cos \theta_W}.
\end{equation}
If we use a conservative lower bound for $M_{W^{\prime}}$ is given by
$M_{W^{\prime}} > 300$ GeV according to
Ref.~\cite{langacker}, the bound for $M_{Z^{\prime}}$ is given by
\begin{equation}
M_{Z^{\prime}} > 340 \ \mbox{GeV}.
\end{equation}
However, for the left-right symmetric case, the stringent bound
\cite{ecker} , $M_{W^{\prime}} > 1.4 -2.5$ TeV, comes from the
$K_L$-$K_S$ mass difference $\Delta m_K$ though this limit strongly
depends on certain theoretical assumptions. In this case the lower
bound for $M_{Z^{\prime}}$ can be as large as 1.6 $-$ 2.9 TeV.

\subsection{Relation between mixing angles and mass ratios}

We can set bounds for the mixing angles $\zeta$ and $\xi$ in terms of
the mass ratios of the gauge bosons. Masso \cite{masso} has derived the
important bound in the charged sector
\begin{equation}
|\zeta| < \frac{M_W^2}{M_{W^{\prime}}^2}
\end{equation}
for the left-right symmetric case with $g_L = g_R$. Langacker and
Sankar \cite{langacker} generalized this bound to the asymmetric case
with $g_L \neq g_R$. We briefly derive the relation and generalize
the bound for the mixing angle in the neutral sector.  

The mixing angle $\zeta$ in the charged sector is given in
Eq.~(\ref{zeta}). For large $v_R$ ($v_R^2 \gg v_L^2, K^2$), the masses
in Eq.~(\ref{zeta}) can be approximated as
\begin{equation}
M_L^2 = \frac{g_L^2}{2} (v_L^2 + K^2) \approx M_W^2, \ M_R^2 =
\frac{g_R^2}{2} (v_R^2 + K^2) \approx \frac{g_R^2}{2} v_R^2 \approx
M_{W^{\prime}}^2.
\label{ml}
\end{equation}
And the mixing is given by $M_{LR}^2 = -g_L g_R |k^* k^{\prime}|$. 
Therefore $\zeta$ can be approximately written as
\begin{equation} 
\zeta \approx -\frac{M_{LR}^2}{M_R^2} \approx \frac{g_L g_R |k^*
k^{\prime}|}{M_{W^{\prime}}^2}.
\end{equation}
Using the Schwarz inequality  $-K^2 < 2|k^* k^{\prime}|<K^2$, we get 
\begin{equation}
-\frac{g_R}{g_L} \frac{M_W^2}{M_{W^{\prime}}^2} < \zeta <
 \frac{g_R}{g_L} \frac{M_W^2}{M_{W^{\prime}}^2}.
\end{equation}
This is the result obtained by Langacker and Sankar, which is a
generalization from the left-right symmetric case.

We can similarly obtain an inequality for the mixing angle $\xi$ in
the neutral sector. The mixing angle $\xi$ is defined in
Eq.~(\ref{xirel}). The masses appearing in $\xi$ can be approximated
as [See Eq.~(\ref{nmass}).] 
\begin{eqnarray}
M_{\tilde{Z}}^2 &=& \frac{g_L^2}{2c_W^2} (K^2 + v_L^2) \approx 
M_Z^2,  \nonumber \\
 M_{\hat{Z}}^2 &=& \frac{1}{2} (g_1^2 + g_R^2)  (v_R^2 + c_R^4K^2 +
s_R^4 v_L^2 )
\approx \frac{g_1^2 + g_R^2}{2} v_R^2 \approx M_{Z^{\prime}}^2.
\end{eqnarray}
And the mixing $M_{\tilde{Z} \hat{Z}}^2$ is given by
\begin{equation}
M_{\tilde{Z} \hat{Z}}^2 = -\frac{1}{2} \frac{g_L}{c_W} g_R c_R  ( K^2
- t_R^2 v_L^2) = -\frac{1}{2} \frac{s_W}{t_R} \frac{g_L^2}{2c_W^2}  
(K^2  - t_R^2 v_L^2). 
\label{mixz}
\end{equation}

Therefore, for large $v_R$, we have
\begin{equation}
\xi \approx \frac{s_W}{t_R} \frac{g_L^2}{2c_W^2} (K^2 - t_R^2
v_L^2) \frac{1}{M_{Z^{\prime}}^2}.
\end{equation}
We use the following relation
\begin{equation}
-t_R^2 ( K^2 + v_L^2) < K^2 -t_R^2 v_L^2 < K^2 + v_L^2
\end{equation}
to have the inequality for $\xi$ as
\begin{equation}
-s_W t_R  \frac{M_Z^2}{M_{Z^{\prime}}^2} < \xi
 < \frac{s_W}{t_R} \frac{M_Z^2}{M_{Z^{\prime}}^2}.
\label{xi}
\end{equation}
This is the generalization of the inequality for the mixing angle
$\xi$ in the neutral sector. 

\subsection{$\rho$ parameter}
The $\rho$ parameter, defined as $\rho \equiv M_W^2/M_Z^2 \cos^2
\theta_W$, can be expressed in terms of mixing angles and gauge boson
mass ratios. Since we have obtained the bounds for the mixing angles,
we can set bounds for the $\rho$ parameter.
In the $SU(2)_L \times SU(2)_R \times U(1)$ model, the masses $M_W$
and $M_Z$ are given in Eqs.~(\ref{wmass}) and (\ref{zmass}) as
\begin{eqnarray}
M_W^2 &=& M_L^2 \cos^2 \zeta +M_R^2 \sin^2 \zeta +M_{LR}^2 \sin
2\zeta, \nonumber \\
M_{Z}^2 &=& M_{\tilde{Z}}^2 \cos^2 \xi + M_{\hat{Z}}^2 \sin^2 \xi
+ M_{\tilde{Z} \hat{Z}}^2 \sin 2\xi,
\label{masses}
\end{eqnarray}
where $\zeta$ and $\xi$ are defined in Eqs.~(\ref{zeta}) and
(\ref{xirel}). 

Since $M_R^2$ and $M_{\hat{Z}}^2$ are of order $v_R^2$, the mixing angles
are small and can be approximately written as 
\begin{equation}
\zeta \approx -\frac{M_{LR}^2}{M_R^2}, \ \ \xi \approx
-\frac{M_{\tilde{Z} \hat{Z}}^2}{M_{\hat{Z}}^2}.
\end{equation}
Using this expression, we can write Eq.~(\ref{masses}) to first order
in the mixing angles as 
\begin{equation}
M_W^2 \approx M_L^2 +\zeta M_{LR}^2, \ \ M_Z^2 \approx M_{\tilde{Z}}^2
+\xi M_{\tilde{Z} \hat{Z}}^2.
\end{equation}
Therefore the $\rho$ parameter can be written as
\begin{eqnarray}
\rho -1 &\approx& \zeta \frac{M_{LR}^2}{M_L^2} -\xi
\frac{M_{\tilde{Z}\hat{Z}}^2}{M_{\tilde{Z}}^2} \nonumber \\
&\approx& -\zeta \frac{g_L g_R |k^* k^{\prime}|}{M_W^2} +\xi \frac{g_L 
g_R c_R (K^2 -t_R^2 v_L^2)/(2c_W)}{M_Z^2}.
\label{rhop}
\end{eqnarray}
The correction to the $\rho$ parameter at tree level is proportional
to the small mixing angles $\zeta$ in the charged sector and $\xi$ in
the neutral sector. 

We can get the bound of $\rho -1$ in Eq.~(\ref{rhop}) using the same
technique in Sec.3.2. Using the inequalities
\begin{equation}
-K^2 < 2|k^* k^{\prime}| < K^2, \ -t_R^2 (K^2
+v_L^2) < K^2 - t_R^2 v_L^2 < K^2 + v_L^2,
\end{equation}
we obtain the relation
\begin{equation}
-( 1+c_R^2 s_R^2) \Bigl( \frac{s_W}{s_R c_R} \Bigr)^2 
\frac{M_Z^2}{M_{Z^{\prime}}^2} < \rho -1 < (1+c_R^4) 
\Bigl( \frac{s_W}{s_R c_R} \Bigr)^2 
\frac{M_Z^2}{M_{Z^{\prime}}^2}.
\label{boundz}
\end{equation}
In deriving Eq.~(\ref{boundz}), we use the tree-level relations $M_W^2
= M_Z^2 \cos^2 \theta_W$, $M_{W^{\prime}}^2 =  M_{Z^{\prime}}^2 \cos^2
\theta_R$ since the corrections to the tree-level values give
higher-order corrections.

Though Eq.~(\ref{boundz}) gives a rough estimate on the bounds of the
$\rho$ parameter, it is not helpful to use this inequality in
numerical estimates since the bounds in Eq.~(\ref{boundz}) may be
overestimated for large $t_R$. Because of this, we do not expect that 
the consideration on $\rho$ from Eq.~(\ref{boundz}) gives a useful
bound for the mass bound of $Z^{\prime}$. Therefore we exclude the
$\rho$ parameter in analyzing the LEP I data.

\section{Phenomenological constraints}
We consider phenomenological constraints on the parameters in the
$SU(2)_L \times SU(2)_R \times U(1)$ model. We employ various
phenomenological inputs such as LEP I data, low-energy neutral-current
data to constrain the parameters $\theta_R$, $\xi$ and
$M_{Z^{\prime}}$. First we obtain the constraints on $\theta_R$ and 
$\xi$ using the LEP I data. With these bounds, we constrain
$\theta_R$, $\xi$ and an additional parameter $M_{Z^{\prime}}$ to
satisfy the low-energy neutral-current data. We present the prediction
for LEP II with the constraints obtained from the combined constraints
from LEP I and low energy data.

We probe all the allowed values of $\xi$, $\theta_R$ and
$M_{Z^{\prime}}$, which satisfy the experimental bounds. However, 
there are some special regions in the  parameter space from
theoretical considerations. First of all, we consider the case 
with $\xi=0$. This can be achieved by setting $K = t_R v_L$. As
can be seen in Eq.~(\ref{mixz}), this is the case where the mixing in
the mass matrix represented by $M_{\tilde{Z}\hat{Z}}^2$ is zero. This
fine tuning seems arbitrary, but it is not. Recall that the scalar
doublet $\chi_L$, which has the VEV $v_L$, is introduced for the
theory to include left-right symmetry easily, but it is not necessary
to attain the desired structure of the theory. Therefore we can vary
$v_L$ along with $\theta_R$ as we want in order to make such a fine
tuning. In this case, the remaining parameters $\theta_R$ and
$M_{Z^{\prime}}^2$ are not constrained by LEP I data since the
corrections are proportional only to the mixing angle $\xi$. Therefore
in this limit, we can evade the precision electroweak LEP I data and
the remaining two parameters are constrained from other experimental
data.

Of course, the value of $v_L$ is not completely arbitrary when we
consider the mechanism why the left-handed neutrinos are so light. For 
example, if the neutrino is of Majorana type, in order for the seesaw
mechanism to work such that the left-handed neutrino mass becomes very
small, $v_L$ cannot be large. However, since there are many variations
in introducing the right-handed neutrino, the possibility for this
fine tuning is still robust. 

The second interesting case is the limit of the left-right symmetry
with $g_L = g_R$. The left-right symmetric model has
been extensively investigated by many authors \cite{moha}. Therefore
we can compare our results with previous analyses. In our version,
this left-right symmetric model puts a definite relation between the
mixing angles $\theta_R$ and $\theta_W$. From Eq.~(\ref{echarge}),
with $g_L =g_R$, the relation between $\theta_R$ and $\theta_W$ is
given by  
\begin{equation}
\sin \theta_R = \tan \theta_W.
\end{equation}
In this case the parameter space is spanned effectively by the two
parameters $\xi$ and $M_{Z^{\prime}}^2$. We will consider this
parameter space also in the following analysis. As it turns out,
the combined results of the LEP I data and the low-energy
neutral-current data raise the lower bound for $M_{Z^{\prime}}$ in the
left-right symmetric theory. This point will be discussed in detail
when we consider the low-energy neutral-current data.

\subsection{Constraints from LEP I data}
In the analysis of the LEP I data, we follow the method employed by
Altarelli et al. \cite{altarelli}. It is equivalent to the analysis
using oblique parameters $S$, $T$ and $U$ \cite{obli}. 
The basic observables in this analysis are the mass ratio
$M_W/M_Z$, the leptonic decay width $\Gamma_{\ell}$, the leptonic
forward-backward asymmetry $A_{FB}^{\ell}$. From these quantities,
we can obtain dynamically significant corrections $\Delta r_W$,
$\Delta \rho$ and $\Delta k$, which contain small effects to be
disentangled. First $\Delta r_W$ is defined from $M_W/M_Z$ by the
relation 
\begin{equation}
\Bigl( 1-\frac{M_W^2}{M_Z^2} \Bigr) \frac{M_W^2}{M_Z^2} = \frac{\pi
\alpha (M_Z)}{\sqrt{2} G_F M_Z^2 (1-\Delta r_W)}.
\label{deltar}
\end{equation}
Here $\alpha(M_Z)$ is fixed to the value 1/128.87.

In order to define $\Delta \rho$ and $\Delta k$, we first write the
coupling of the $Z$ particle to charged leptons in the form
$\overline{\ell} \gamma^{\mu} (g_V -g_A \gamma_5) \ell$. Then the
physical quantities $\Gamma_{\ell}$ and $A_{FB}^{\ell}$ can be
parameterized by the effective vector and axial-vector couplings $g_V$
and $g_A$ as  
\begin{equation}
\Gamma_{\ell} = \frac{G_F M_Z^3}{6\pi \sqrt{2}} (g_V^2 + g_A^2)
\Bigl( 1+\frac{3\alpha}{4\pi} \Bigr), \ \  A_{FB}^{\ell}(\sqrt{s}=M_Z) 
= \frac{3g_V^2 g_A^2}{(g_V^2 + g_A^2)^2}. 
\end{equation}
Here $\Gamma_{\ell}$ stands for the inclusive partial width $\Gamma (Z 
\rightarrow \ell \ell +\mbox{photons})$. $\Delta \rho$ and $\Delta k$
are defined as 
\begin{equation} 
g_A = -\frac{1}{2} \Bigl(1+ \frac{\Delta
\rho}{2} \Bigr), \ \
\frac{g_V}{g_A} = 1-4(1+\Delta k) s_W^2,
\label{defrk}
\end{equation}
where $s_W^2$ is evaluated at tree level, given by
\begin{equation}
s_W^2 c_W^2 = \frac{\pi \alpha(M_Z)}{\sqrt{2} G_F M_Z^2},
\end{equation}
with $c_W^2 = 1-s_W^2$. ($s_W^2 =0.231184$ for $M_Z=91.187$ GeV.) 

We express $\Delta \rho$, $\Delta r_W$ and $\Delta k$ in terms of
the following combinations:
\begin{eqnarray}
\epsilon_{1} &=& \Delta \rho, \nonumber \\
\epsilon_{2} &=& c_W^2 \Delta \rho +\frac{s_W^2 \Delta r_W}{c_W^2
-s_W^2} -2s_W^2 \Delta k, \nonumber \\
\epsilon_{3} &=& c_W^2 \Delta \rho + (c_W^2 -s_W^2) \Delta k.
\end{eqnarray}
These variables $\epsilon_{i}$ are used in analyzing new physics
effects beyond the standard model. In our case, the additional
corrections are expressed in terms of the mixing angle $\xi$,
$\theta_R$. Note that the ratio of the masses $M_Z^2/M_{Z^{\prime}}^2$
does not enter the analysis for LEP I data since the LEP I experiment
is performed at the $Z$ peak.

Note that only the parameter $\epsilon_{2}$ depends on $\Delta r_W$,
hence on $M_W^2/M_Z^2$ as shown in Eq.~(\ref{deltar}). The ratio
$M_W^2/M_Z^2$ is related to the $\rho$ parameter and the approximate
relation for the $\rho$ parameter is given in Eq.~(\ref{rhop}). There
are two kinds of uncertainties in this expression. First, it depends
on the information of the charged sector which needs an independent
analysis. Secondly, it is difficult to express the term belonging to
the neutral sector in terms of the parameters $\xi$, $\theta_R$ and
$M_Z^{\prime}$. For these reasons, we avoid using $\epsilon_2$, and we
will consider constraints only in the $(\epsilon_{1},\epsilon_{3})$
parameter space.   

\begin{figure}[ht]
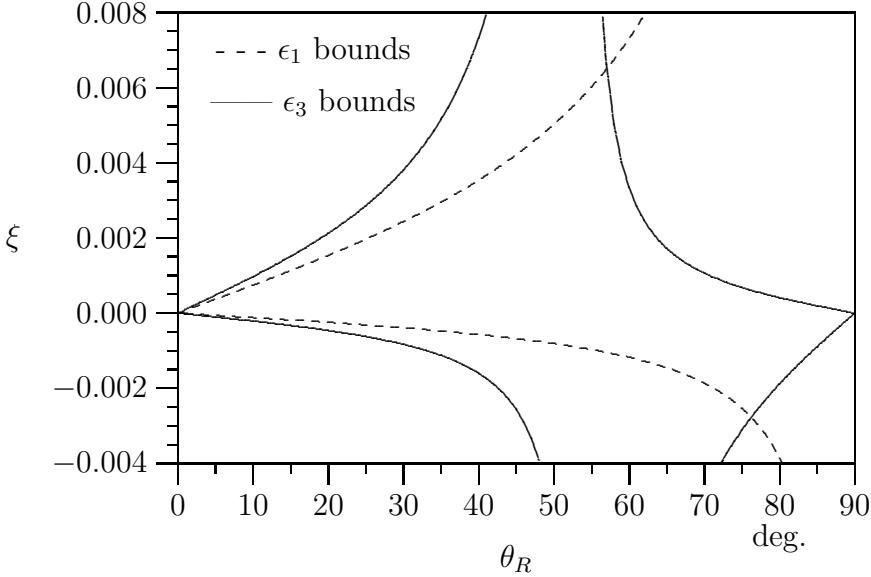

\vspace{0.4in}
\beginpicture
\setcoordinatesystem units <0.1cm,0.1cm> 
\setplotarea x from 0 to 90, y from -20 to 40
\axis bottom label {$\theta_R$} ticks
length <4pt> unlabeled from  0 to 90 by 5.0
length <8pt> width <.4pt> withvalues 0 10 20 30 40 50 60 70 80 90 /
at 0.0 10.0 20.0 30.0 40.0 50.0 60.0 70.0 80.0 90.0 / /
\axis top  ticks /
\axis left label {$\xi$} ticks
length <4pt> unlabeled from -20 to 40 by 2.5 
length <8pt> width <.4pt> withvalues $-$0.004 $-$0.002 0.000 0.002
0.004 0.006 0.008   /
at -20.0 -10.0 0.0 10.0 20.0 30.0 40.0 / /
\axis right ticks /
\put {- - -   $\epsilon_1$ bounds} at 18 35
\put {------ $\epsilon_3$ bounds} at 18 28
\put {deg.} at 80 -30
\setdashes <3pt>
\setlinear
\plot   
  1.0     -.06
  2.0     -.12
  3.0     -.18
  4.0     -.24
  5.0     -.30
  6.0     -.36
  7.0     -.42
  8.0     -.48
  9.0     -.54
 10.0     -.60
 11.0     -.66
 12.0     -.72
 13.0     -.78
 14.0     -.85
 15.0     -.91
 16.0     -.97
 17.0    -1.04
 18.0    -1.10
 19.0    -1.17
 20.0    -1.23
 21.0    -1.30
 22.0    -1.37
 23.0    -1.44
 24.0    -1.51
 25.0    -1.58
 26.0    -1.65
 27.0    -1.73
 28.0    -1.80
 29.0    -1.88
 30.0    -1.96
 31.0    -2.04
 32.0    -2.12
 33.0    -2.20
 34.0    -2.29
 35.0    -2.37
 36.0    -2.46
 37.0    -2.55
 38.0    -2.65
 39.0    -2.75
 40.0    -2.84
 41.0    -2.95
 42.0    -3.05
 43.0    -3.16
 44.0    -3.27
 45.0    -3.39
 46.0    -3.51
 47.0    -3.64
 48.0    -3.76
 49.0    -3.90
 50.0    -4.04
 51.0    -4.19
 52.0    -4.34
 53.0    -4.50
 54.0    -4.67
 55.0    -4.84
 56.0    -5.03
 57.0    -5.22
 58.0    -5.43
 59.0    -5.64
 60.0    -5.87
 61.0    -6.12
 62.0    -6.38
 63.0    -6.65
 64.0    -6.95
 65.0    -7.27
 66.0    -7.61
 67.0    -7.99
 68.0    -8.39
 69.0    -8.83
 70.0    -9.31
 71.0    -9.85
 72.0   -10.43
 73.0   -11.09
 74.0   -11.82
 75.0   -12.65
 76.0   -13.60
 77.0   -14.68
 78.0   -15.95
 79.0   -17.44
 80.0   -19.23
 80.1   -19.42
 80.2   -19.63
 80.3   -19.83 /
\plot
   .0      .00
  1.0      .37
  2.0      .74
  3.0     1.10
  4.0     1.47
  5.0     1.84
  6.0     2.21
  7.0     2.59
  8.0     2.96
  9.0     3.34
 10.0     3.71
 11.0     4.10
 12.0     4.48
 13.0     4.86
 14.0     5.25
 15.0     5.65
 16.0     6.04
 17.0     6.44
 18.0     6.85
 19.0     7.25
 20.0     7.67
 21.0     8.09
 22.0     8.51
 23.0     8.94
 24.0     9.38
 25.0     9.82
 26.0    10.28
 27.0    10.73
 28.0    11.20
 29.0    11.68
 30.0    12.16
 31.0    12.66
 32.0    13.16
 33.0    13.68
 34.0    14.21
 35.0    14.75
 36.0    15.31
 37.0    15.88
 38.0    16.46
 39.0    17.06
 40.0    17.68
 41.0    18.31
 42.0    18.97
 43.0    19.65
 44.0    20.34
 45.0    21.07
 46.0    21.82
 47.0    22.59
 48.0    23.40
 49.0    24.24
 50.0    25.11
 51.0    26.02
 52.0    26.97
 53.0    27.96
 54.0    29.00
 55.0    30.09
 56.0    31.23
 57.0    32.44
 58.0    33.72
 59.0    35.06
 60.0    36.49
 61.0    38.01
 62.0    39.62 /
\setsolid
\plot
   .1     -.01
   .2     -.02
   .3     -.03
   .4     -.04
   .5     -.05
   .6     -.06
   .7     -.07
   .8     -.08
   .9     -.09
  1.0     -.10
  2.0     -.21
  3.0     -.31
  4.0     -.41
  5.0     -.52
  6.0     -.63
  7.0     -.73
  8.0     -.84
  9.0     -.95
 10.0    -1.06
 11.0    -1.17
 12.0    -1.29
 13.0    -1.41
 14.0    -1.53
 15.0    -1.65
 16.0    -1.77
 17.0    -1.90
 18.0    -2.04
 19.0    -2.18
 20.0    -2.32
 21.0    -2.47
 22.0    -2.62
 23.0    -2.78
 24.0    -2.95
 25.0    -3.12
 26.0    -3.31
 27.0    -3.50
 28.0    -3.71
 29.0    -3.93
 30.0    -4.16
 31.0    -4.41
 32.0    -4.68
 33.0    -4.97
 34.0    -5.28
 35.0    -5.62
 36.0    -6.00
 37.0    -6.42
 38.0    -6.88
 39.0    -7.40
 40.0    -7.99
 41.0    -8.66
 42.0    -9.44
 43.0   -10.36
 44.0   -11.45
 45.0   -12.79
 46.0   -14.46
 47.0   -16.61
 48.0   -19.50 /

\plot
 56.5    39.34
 56.6    37.93
 56.7    36.61
 56.8    35.38
 56.9    34.23
 57.0    33.15
 58.0    25.12
 59.0    20.13
 60.0    16.72
 61.0    14.23
 62.0    12.34
 63.0    10.84
 64.0     9.62
 65.0     8.61
 66.0     7.76
 67.0     7.02
 68.0     6.38
 69.0     5.82
 70.0     5.31
 71.0     4.86
 72.0     4.45
 73.0     4.07
 74.0     3.72
 75.0     3.40
 76.0     3.10
 77.0     2.82
 78.0     2.55
 79.0     2.30
 80.0     2.06
 81.0     1.83
 82.0     1.61
 83.0     1.39
 84.0     1.18
 85.0      .98
 86.0      .78
 87.0      .58
 88.0      .39
 89.0      .19
 90.0      .00 /
\plot
  1.0        .47
  2.0        .95
  3.0       1.42
  4.0       1.90
  5.0       2.38
  6.0       2.86
  7.0       3.35
  8.0       3.85
  9.0       4.35
 10.0       4.85
 11.0       5.37
 12.0       5.90
 13.0       6.43
 14.0       6.98
 15.0       7.54
 16.0       8.12
 17.0       8.71
 18.0       9.33
 19.0       9.96
 20.0      10.61
 21.0      11.29
 22.0      11.99
 23.0      12.72
 24.0      13.49
 25.0      14.29
 26.0      15.14
 27.0      16.03
 28.0      16.97
 29.0      17.97
 30.0      19.04
 31.0      20.18
 32.0      21.41
 33.0      22.73
 34.0      24.17
 35.0      25.74
 36.0      27.46
 37.0      29.36
 38.0      31.48
 39.0      33.86
 40.0      36.55
 41.0      39.63 /
\plot
 72.2     -20.00
 72.3     -19.83
 72.4     -19.65
 72.5     -19.48
 73.0     -18.64
 74.0     -17.05
 75.0     -15.57
 76.0     -14.20
 77.0     -12.90
 78.0     -11.69
 79.0     -10.53
 80.0      -9.42
 81.0      -8.37
 82.0      -7.35
 83.0      -6.36
 84.0      -5.40
 85.0      -4.47
 86.0      -3.55
 87.0      -2.65
 88.0      -1.76
 89.0       -.88
 90.0        .00 /
\endpicture
\caption{Bounds on $\xi$ and $\theta_R$ from $\epsilon_1$ and
$\epsilon_3$ using the LEP I data at 95\% CL. Solid curves are the
bounds from $\epsilon_3$ and the region enclosed by four curves is
allowed. Dashed curves are the bounds from $\epsilon_1$ and the region
between two dashed curves is allowed.}  
\label{fig1} 
\end{figure}

In terms of $\xi$ and $t_R$, the parameters $\epsilon_{1}$ and
$\epsilon_{3}$ are written as 
\begin{equation}
\epsilon_{1} = \epsilon_1^{\mathrm{SM}} +\Delta \rho_{LR}, \ 
\epsilon_{3} = \epsilon_3^{\mathrm{SM}} + c_W^2 \Delta \rho_{LR} +
(c_W^2 -s_W^2) \Delta k_{LR},
\end{equation}
where 
\begin{equation}
\Delta \rho_{LR} = -\frac{2\xi s_W}{t_R}, \ \ \Delta k_{LR} = \xi s_W
\Bigl[\frac{1}{t_R} +\frac{1}{2s_W^2} \bigl( t_R -\frac{1}{t_R} \bigr)
\Bigr].
\end{equation}
The quantities $\epsilon_1^{\mathrm{SM}}$, $\epsilon_3^{\mathrm{SM}}$
are the contributions from the standard model. In the numerical
analysis, we use the values of $\epsilon_1^{\mathrm{SM}}$ and
$\epsilon_3^{\mathrm{SM}}$ including the electroweak radiative
corrections as encoded in ZFITTER \cite{zfitter}. The experimental
values for $\epsilon_1$ and $\epsilon_3$ are given by
\begin{equation}
\epsilon_1 = (2.6 - 5.3) \times 10^{-3}, \ \ \epsilon_3 = (1.0 - 4.5)
\times 10^{-3}, 
\end{equation}
which can be calculated from Ref.~\cite{lepewwg}

Using the variables $\epsilon_1$ and $\epsilon_3$, we can put
constraints on the parameters $\xi$ and $\theta_R$. The constraints on
$\xi$ and $\theta_R$ are shown in Fig.~\ref{fig1}. The solid curves
represent the constraint by $\epsilon_3$ and the dashed curves
represent the constraint by $\epsilon_1$. From the overlapping region
of these two bounds in Fig.~\ref{fig1}, we obtain
\begin{equation}
-0.0028 < \xi < 0.0065,
\end{equation}
while there is little constraint on $\theta_R$.

Note that these bounds are for all possible values of $\theta_R$. We
can easily get bounds for $\xi$ at some specific values of
$\theta_R$. For example, the left-right symmetric model corresponds to
$\theta_R \approx 33^{\circ}$ ($\sin \theta_R = \tan \theta_W$). In
this case, the bound for $\xi$ is $-0.0005 < \xi < 0.0026$. As shown
in Fig.~\ref{fig1}, the characteristic of the relation between
$\theta_R$ and $\xi$ is that positive (negative) values of $\xi$ are
preferred for small (large) $\theta_R$.

\begin{figure}[ht]
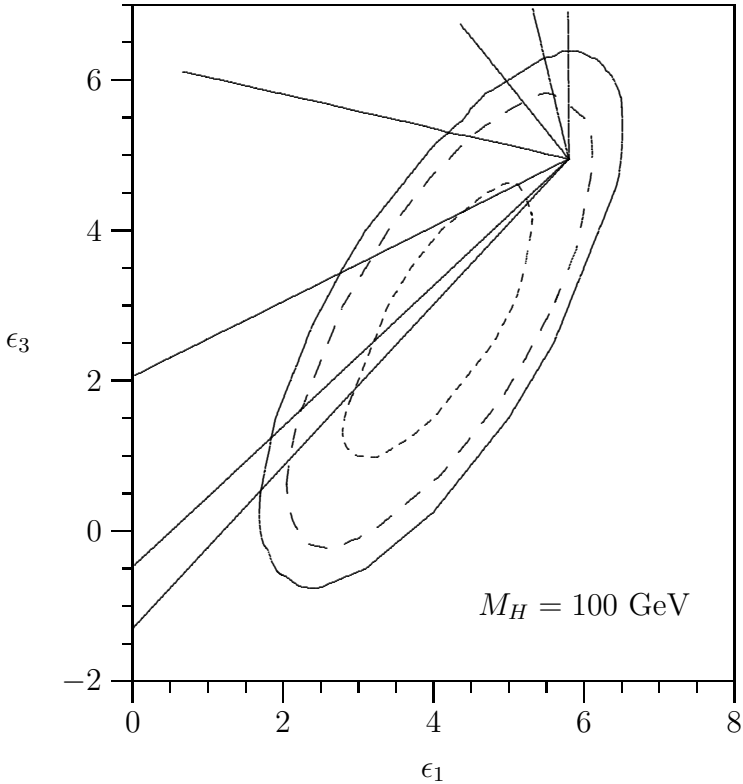

\vspace{0.4in}
\beginpicture
\setcoordinatesystem units <0.1cm,0.1cm> 
\setplotarea x from 0 to 80, y from -20 to 70
\axis bottom label {$\epsilon_1$} ticks
length <4pt> unlabeled from  0 to 80 by 5.0
length <8pt> width <.4pt> withvalues 0 2 4 6 8 /
at 0.0 20.0 40.0 60.0 80.0 / /
\axis top  ticks /
\axis left label {$\epsilon_3$} ticks
length <4pt> unlabeled from -20 to 70 by 5.0 
length <8pt> width <.4pt> withvalues $-$2 0 2 4 6   /
at -20.0 0.0 20.0 40.0 60.0  / /
\axis right ticks /
\setdashes <3pt>
\setquadratic
\plot
29.0  17.5 30.0  20.0 31.0  22.5 32.0  25.0 33.0  27.5 34.0  30.0 35.0
31.67 36.0  33.333 37.0  35.00 38.0  36.25 39.0  37.50 40.0  38.75
41.0  40.0 42.0  40.833 43.0  41.667 44.0  42.50 45.0  43.333 46.0
44.167 47.0  45.0 48.0  45.4 49.0  46.0 50.0  46.3 51.0  46.1 52.0
45.0 53.0  42.3 53.0  40.0 52.0  34.5 51.0  31.0 50.0  29.0 49.0  27.0
48.0  25.0 47.0  23.75 46.0  22.5 45.0  21.25 44.0  20.0 43.0  18.75
42.0  17.5 41.0  16.2 40.0  15.0 39.0  14.167 38.0  13.33 37.0  12.5
36.0  11.875 35.0  11.25 34.0  10.625 33.0  10.0 32.0  9.8 31.0  9.9
30.0  10.0 28.0  12.5  29.0 17.5 /
\setdashes <7pt>
\plot
20.5  6.25 21.0   10.0 22.0   15.0 23.0   17.5 24.0   20.0 25.0   22.5
26.0   25.0 27.0   27.5 28.0   30.0 29.0   31.667 30.0   33.333 31.0
35.0 32.0   36.667 33.0   38.333 34.0   40.0 35.0   41.25 36.0   42.5
37.0   43.75 38.0   45.0 39.0   46.25 40.0   47.5 41.0   48.333 42.0
49.3 43.0   50.5 44.0   51.5 45.0   52.50 46.0   53.2 47.0   54.0 48.0
54.5 49.0   55.2 50.0   56.0 51.0   56.5 52.0   57.0 53.0   57.5 54.0
58.0 55.0   58.2 56.0   58.0 57.0   57.5 58.0   56.667 59.0   55.833
60.0   55.0 61.0   52.5 61.0   47.5 60.0   42.5 59.0   37.5 57.0
32.5 56.0   30.0 55.0   28.125 54.0   26.25 53.0   24.375 52.0   22.5
51.0   20.833 50.0   19.167 49.0   17.5 48.0   16.25 47.0   15.0 46.0
13.75 45.0   12.5 44.0   11.25 43.0   10.0 42.0   8.75 41.0   7.50
40.0   6.667 39.0   5.833 38.0   5.0 37.0   4.167 36.0   3.333 35.0
2.5 34.0   1.667 33.0   0.833 32.0   0.00 31.0   -0.4 30.0   -1.0 29.0
-1.3 28.0   -1.9 27.0   -2.1 26.0   -2.25 25.0   -2.1 24.0   -1.9 23.0
-1.4 22.0   0.1 21.0   2.5  20.5   6.25 /
\setsolid
\plot 17.0  5.0 18.0  10.0 19.0  15.0 20.0  17.5 21.0  20.0 22.0  22.5
23.0  25.0 24.0  27.5 25.0  29.375 26.0  31.25 27.0  33.125 28.0  35.0
29.0  36.667 30.0  38.333 31.0  40.0 32.0  41.25 33.0  42.5 34.0
43.75 35.0  45.0 36.0  46.25 37.0  47.5 38.0  48.75 39.0  50.0 40.0
51.2 41.0  52.2 42.0  53. 43.0  54. 44.0  54.8 45.0  56.2 46.0  57.2
47.0  58.3 48.0  58.75 49.0  59.375 50.0  60.0 51.0  60.625 52.0
61.25 53.0  61.875 54.0  62.5 55.0  63. 56.0  63.2 57.0  63.8 58.0
63.9 59.0  63.9 60.0  63.5 61.0  63.0 62.0  62.5 63.0  61.25 64.0
60.0 65.0  57.5 65.0  50.0 64.0  45.0 63.0  42.5 62.0  40.0 61.0  37.5
60.0  35.0 59.0  32.5 58.0  30.0 57.0  27.5 56.0  25.0 55.0  23.333
54.0  21.667 53.0  20.0 52.0  18.333 51.0  16.667 50.0  15.0 49.0
13.75 48.0  12.5 47.0  11.25 46.0  10.0 45.0  8.75 44.0  7.5 43.0
6.25 42.0  5.0 41.0  3.75 40.0  2.5 39.0  1.667 38.0  0.833 37.0  0.0
36.0 -0.833 35.0 -1.667 34.0 -2.5 33.0 -3.333 32.0 -4.167 31.0 -5.0
30.0 -5.417 29.0 -5.833 28.0 -6.25 27.0 -6.667 26.0 -7.073 25.0 -7.5
24.0 -7.6 23.0 -7.5 22.0 -7.3 21.0 -6.5 20.0 -5.833 19.0 -5.0 18.0
-2.5 17.0  0.0  17.0 5.0 /
\setsolid
\setlinear
\plot  58.0  49.5  10.0 -2.20  0.0 -13.0 /
\plot   58.0  49.5  48.4  40.5  38.8  31.5  29.2  22.5  19.6  13.5
10.0  4.50   0.30 -4.50  /
\plot   58.0  49.5  53.2  47.1  48.4  44.7  43.6  42.3  38.8  39.9
34.0  37.5  29.2  35.1  24.4  32.7  19.6  30.3  14.8  27.9  9.90  25.5
5.10  23.1  0.30  20.7 /
\plot   58.0  49.5  54.8  50.2  51.6  51.0  48.4  51.7  45.2  52.4
42.01  53.0   38.8  53.9  35.6  54.6  32.4  55.3  29.2  56.0  26.0
56.8   22.8  57.5  19.6  58.2  16.4  58.9  13.1  59.7  9.90  60.4
6.70  61.1 /
\plot   58.0  49.5  55.6  52.5  53.2  55.5  50.8  58.5  48.4  61.5
46.0  64.5  43.6  67.4 /
\plot   58.0  49.5  56.4  56.2  54.8  62.8  53.2  69.5 /
\plot   58.0  49.5  57.9  69.0 /
\put{$M_H = 100$ GeV} at 60 -10 
\endpicture
\caption{Behavior of $(\epsilon_1, \epsilon_3)$ as $\xi$ and
$\theta_R$ are varied. The ellipses are obtained from experimental
data. The solid curve corresponds to  95\% CL, the long-dashed curve
to 90\% CL and the short-dashed curve to  1$\sigma$ level. The
straight lines represent the behavior of ($\epsilon_1, \epsilon_3$)
with $t_R =$ 0.1, 0.5, 1.0, 1.5, 2.0, 3.0 and 
20.0 respectively, starting from the lower left line to the clockwise
direction. Positive values of $\xi$ only are shown.}
\label{fig2} 
\end{figure}

We can also consider the behavior of $(\epsilon_1,\epsilon_3)$ as
we vary $\xi$ and $\theta_R$. It is shown in Fig.~\ref{fig2} for the
Higgs mass $M_H=100$ GeV. The ellipses are bounds from LEP I
data. They represent 95\% CL, 90\% CL, and $1\sigma$-level estimates
respectively as we go inside. The converging point corresponds to the
standard model value. The straight lines show the behavior of
$\epsilon_1$ and $\epsilon_3$ as we vary $\theta_R$. As we increase
$\theta_R$, the lines move in the clockwise direction. We show the
behavior for positive values of $\xi$ only. As we increase $\xi$, the
values of $\epsilon_1$ and $\epsilon_3$ deviate further away from the
standard model value. For negative values of $\xi$, the direction is
reversed. As we vary the Higgs mass $M_H$ from 100 GeV to 1 TeV, this
qualitative feature does not change though the point representing the
standard model moves downward a little bit. In order for the point
$(\epsilon_{1},\epsilon_{3})$ to approach the central region of the
data, $\xi$ should be negative for large values of $\theta_R$ or $\xi$
should be positive for small values of $\theta_R$. We can understand
this behavior clearly in Fig.~\ref{fig1}, as already pointed out.
 
In the special case for $\xi =0$, there is no deviation from the
standard model since all the corrections are proportional to
$\xi$. Therefore with this fine tuning ($K = t_R v_L$), the
standard-model prediction remains intact. The constraint on
$\xi$, with $\theta_R$ varied, from LEP I will be combined with the
low-energy neutral-current data to further constrain all the three
parameters $\xi$, $\theta_R$ and $M_{Z^{\prime}}$ in Sec. 4.2.

\subsection{Constraints from low-energy neutral-current data}
Now we consider the low-energy neutral-current interactions such
as $\nu e\rightarrow \nu e$,  $\nu N$ scattering, and $e_{L,R}N
\rightarrow e_{L,R} X$ in which both $Z$ and $Z^{\prime}$ can
participate. As far as the neutral-current interactions
involving neutrinos are concerned, note that there is one striking
difference in the $SU(2)_L \times SU(2)_R \times U(1)$ model compared
to the standard model. It is the existence of the right-handed 
neutrino. There are a few possibilities to include the right-handed
neutrino, such as a heavy Dirac neutrino, medium-mass neutrino, light
neutrino, or heavy Majorana neutrino. The Majorana-type neutrino
provides an interesting way to make the left-handed neutrino light via
the seesaw mechanism \cite{moha}. 

We include both left-handed and right-handed neutrinos in the
analysis. However, if we keep the corrections to first order in $\xi$
or $M_Z^2/M_{Z^{\prime}}^2$, the right-handed neutrino does not
contribute to the physical observables irrespective of the types of
right-handed neutrinos. In order to illustrate this point, consider
the process $\nu_e e\rightarrow \nu_e e$. The effective Hamiltonian for
this process is of the form 
\begin{eqnarray}
H_{\mathrm{eff}} &=& \frac{G_F}{\sqrt{2}} \overline{\nu} \Bigl
( g_L^{\nu} \gamma^{\mu} (1-\gamma_5) + g_R^{\nu} \gamma^{\mu}
(1+\gamma_5) \Bigr) \nu \nonumber \\
&&\times \overline{e} \Bigl( g_L^e \gamma^{\mu}
(1-\gamma_5) + g_R^e \gamma^{\mu} 
(1+\gamma_5) \Bigr) e.
\label{effham}
\end{eqnarray}

The value of $g_R^{\nu}$ in the standard model is zero and  it starts
from the first order in $\xi$ or $M_Z^2/M_{Z^{\prime}}^2$. In the
matrix element squared, the nonzero contribution has an even number of
left-handed and right-handed currents due to helicity conservation.
The matrix element squared for any process from Eq.~(\ref{effham})
depends on $g_R^{\nu 2}$, hence of second order in $\xi$ or
$M_Z^2/M_{Z^{\prime}}^2$. Therefore to first order in $\xi$ or
$M_Z^2/M_{Z^{\prime}}^2$, we can safely disregard the contribution
from the right-handed neutrino.

For the scattering $\nu e \rightarrow \nu e$, only the $Z$ and
$Z^{\prime}$ particles participate in the interactions. The relevant
effective Hamiltonian at low energy can be written in the
form  
\begin{equation}
H^{\nu e} = \frac{G_F}{\sqrt{2}} \overline{\nu}
\gamma^{\mu}  (1-\gamma_5) \nu \overline{e} \gamma_{\mu} (g_V^{\nu e}
- g_A^{\nu e} \gamma_5) e,
\end{equation}
neglecting the contribution from the right-handed neutrino as
discussed above. We can obtain $g_V^{\nu e}$ and $g_A^{\nu e}$ to
first order in the small mixing angle $\xi$ and the mass ratio
$M_Z^2/M_{Z^{\prime}}^2$. Note that $\xi$ is of order
$O(M_Z^2/M_{Z^{\prime}}^2)$, therefore we have to keep both terms in
order to be consistent. The coupling constants $g_V^{\nu e}$ and
$g_A^{\nu e}$ are written as 
\begin{eqnarray}
g_V^{\nu e} &=& -\frac{1}{2} +2s_W^2 +\xi s_W \Bigl[ \Bigl(\frac{1}{2}
+ 2s_W^2\Bigr) t_R -\frac{1}{2t_R} \Bigr] +
\frac{M_Z^2}{M_{Z^{\prime}}^2} s_W^2 \Bigl(t_R^2 -\frac{1}{2} \Bigr),
\nonumber \\ 
g_A^{\nu e} &=& -\frac{1}{2} -\frac{\xi s_W}{2} \Bigl(t_R
-\frac{1}{t_R} \Bigr)
+\frac{1}{2}\frac{M_Z^2}{M_{Z^{\prime}}^2} s_W^2.
\label{nue}
\end{eqnarray}
In Eq.~(\ref{nue}), those terms independent of $\xi$ and
$M_Z^2/M_{Z^{\prime}}^2$ are the values from the standard model.

For neutrino-hadron scattering, the relevant effective Hamiltonian can
be written as
\begin{eqnarray}
H^{\nu N} &=& \frac{G_F}{\sqrt{2}} \overline{\nu}
\gamma^{\mu} (1-\gamma_5) \nu \nonumber \\
&\times& \sum_i \Bigl[ \epsilon_L(i) \overline{q}_i \gamma_{\mu}
(1-\gamma_5) q_i + \epsilon_R (i) \overline{q}_i \gamma_{\mu} (1+
\gamma_5) q_i \Bigr],
\end{eqnarray}
where $\epsilon_{L,R} (i)$ are given by
\begin{eqnarray}
\epsilon_L (u) &=& \frac{1}{2} -\frac{2}{3} s_W^2 + \frac{s_W t_R}{3}
\Bigl[ \xi (1-2 s_W^2 ) -\frac{1}{2}
\frac{M_Z^2}{M_{Z^{\prime}}^2} s_W t_R \Bigr], \nonumber \\
\epsilon_L (d) &=& -\frac{1}{2} + \frac{1}{3} s_W^2 + \frac{s_W
t_R}{3} \Bigl[ \xi (-2+s_W^2) -\frac{1}{2}
\frac{M_Z^2}{M_{Z^{\prime}}^2} s_W t_R\Bigr], \nonumber \\ 
\epsilon_R (u) &=& -\frac{2}{3} s_W^2 + \frac{s_W t_R}{3} \Bigl[\xi 
(-\frac{1}{2} -2 s_W^2 + \frac{3}{2t_R^2}) -
\frac{1}{2} \frac{M_Z^2}{M_{Z^{\prime}}^2}  s_W t_R \Bigl( 1
-\frac{3}{t_R^2} \Bigl) \Bigr], \nonumber \\ 
\epsilon_R (d) &=& \frac{1}{3} s_W^2 + \frac{s_W t_R}{3} \Bigl[ \xi
\Bigl( -\frac{1}{2} + s_W^2 -\frac{3}{2t_R^2} \Bigr) - \frac{1}{2}
\frac{M_Z^2}{M_{Z^{\prime}}^2} s_W t_R \Bigl( 1+\frac{3}{t_R^2} \Bigr)
\Bigr]. 
\label{eps}
\end{eqnarray}

For electron-hadron scattering such as $e_{L,R} N \rightarrow eX$
performed in the SLAC polarized electron experiment, the
parity-violating Hamiltonian  can be written as
\begin{equation}
H^{eN} = -\frac{G_F}{\sqrt{2}} \sum_i \Bigl[ C_{1i} \overline{e}
\gamma^{\mu} \gamma_5 e \overline{q}_i \gamma_{\mu} q_i + C_{2i}
\overline{e} \gamma^{\mu} e \overline{q}_i \gamma_{\mu} \gamma_5 q_i
\Bigr]. 
\label{hen}
\end{equation}
The coefficients $C_{1,2i}$ are given by
\begin{eqnarray}
C_{1u} &=& -\frac{1}{2} +\frac{4}{3} s_W^2 + \xi s_W
\Bigl(\frac{t_R}{3} -\frac{4}{3} \frac{s_W^2}{t_R} \Bigr) -
\frac{M_Z^2}{M_{Z^{\prime}}^2} s_W^2  
\Bigl(\frac{1}{3} - \frac{1}{2t_R^2} \Bigr),
\nonumber \\ 
C_{2u}&=& -\frac{1}{2} + 2s_W^2 + \xi s_W \Bigl(t_R-\frac{2s_W^2}{t_R}
\Bigr) - \frac{M_Z^2}{M_{Z^{\prime}}^2} s_W^2 \Bigl(1-\frac{1}{2t_R^2}
\Bigr), \nonumber
\\
C_{1d}&=& \frac{1}{2} -\frac{2}{3} s_W^2 + \frac{\xi s_W}{3} \Bigl(t_R
+ \frac{2s_W^2}{t_R} \Bigr) - \frac{M_Z^2}{M_{Z^{\prime}}^2} s_W^2
\Bigl(\frac{1}{3} + \frac{1}{2t_R^2} \Bigr), \nonumber \\
C_{2d} &=& \frac{1}{2} -2s_W^2 - \xi s_W \Bigl(t_R -
\frac{2s_W^2}{t_R} \Bigr) + \frac{M_Z^2}{M_{Z^{\prime}}^2} s_W^2
\Bigl(1-\frac{1}{2t_R^2} \Bigr). 
\label{cs}
\end{eqnarray}

\begin{table}[ht]
\vspace{0.4in}
\begin{tabular}{ccc} \hline 
Quantity&Experiments& SM prediction \\ \hline 
$\epsilon_L (u)$ & 0.328$\pm$0.016 & 0.3461$\pm$0.0002 \\
$\epsilon_L (d)$ & $-$0.440$\pm$0.011 & $-$0.4292$\pm$0.0002 \\
$\epsilon_R (u)$ & $-$0.179$\pm$0.013& $-$0.1548$\pm$0.0001 \\
$\epsilon_R (d)$ & $-0.027^{+0.077}_{-0.048}$& 0.0775$\pm$0.0001 \\
\hline
$g_V^{\nu e}$ & $-$0.041$\pm$0.015& $-$0.0395 $\pm$0.0005 \\ 
$g_A^{\nu e}$ & $-$0.507$\pm$0.014 & $-$0.5064$\pm$0.0002\\ \hline
$C_{1u}$ & $-$0.216$\pm$0.046 & $-$0.1885$\pm$0.0003\\
$C_{1d}$ & 0.361$\pm$0.041 & 0.3412$\pm$0.0002 \\ 
$C_{2u}-\frac{1}{2}C_{2d}$& $-$0.03$\pm$0.12 & $-$0.0488$\pm$0.0008\\
\hline
$Q_W$ & $-72.41\pm 1.05$ & $-73.20 \pm 0.13$ \\ \hline
\end{tabular}
\vspace{0.2in}
\caption{Values of the model-independent neutral-current parameters,
compared with the standard model predictions for $M_Z = 91.1867$ GeV
($M_H = M_Z$) \cite{pdg}.} 
\end{table}

The atomic parity violation can be described by the Hamiltonian in
Eq.~(\ref{hen}). The weak charge of an atom is defined as
\begin{equation}
Q_W = -2 \Bigl[ C_{1u} (2Z+N) + C_{1d} (Z+2N) \Bigr],
\end{equation}
where $Z$($N$) is the number of protons(neutrons) in the atom. For
${}^{133}_{55} \mbox{Cs}$  atom, $Z=55$, $N=78$, the correction to the
weak charge is given by 
\begin{eqnarray}
\Delta Q_W &\equiv& Q_W - Q_W^{\mathrm{SM}} 
= -2 \Bigl[ \Delta C_{1u} (2Z+N) + \Delta C_{1d} (Z+2N) \Bigl]
\nonumber \\
&=& s_W \Bigl[ \xi t_R \Bigl( -266 + 220 \frac{s_W^2}{t_R^2} \Bigr) +
\frac{M_Z^2}{M_{Z^{\prime}}^2} s_W \Bigl( 266 + \frac{23}{t_R^2}
\Bigr) \Bigr]. 
\end{eqnarray}
The recent measurement and the analysis of the weak charge for the Cs
atom  gives the value \cite{apv_exp}
\begin{equation}
\Delta Q_W  = 0.79 \pm 1.06.
\end{equation}

All the experimental values for the observables and the
standard model prediction are tabulated in Table~1. We use the ten
physical observables listed in Table~1 to constrain $\xi$, $\theta_R$
and $M_{Z^{\prime}}$ at 95\% CL. As shown in Eqs.~(\ref{nue}),
(\ref{eps}) and (\ref{cs}),  since there are two kinds of terms
proportional to $\xi$ and $M_Z^2/M_{Z^{\prime}}^2$, and since $\xi$
can be either positive or negative, the contribution of these terms
may be partially cancelled. If the relative sign of these two terms
in a quantity is opposite, both parameters $|\xi|$ and
$M_Z^2/M_{Z^{\prime}}^2$ can be large without exceeding experimental
bounds. Due to this fact, the bounds obtained by the low-energy
neutral-current data alone are not useful. In order to obtain useful
bounds, we first constrain the parameters $\xi$ and $\theta_R$ from
LEP I data, and then we look for constraints on $\xi$, $\theta_R$ and
$M_{Z^{\prime}}$ satisfying the low-energy neutral-current data.

\begin{figure}[ht]
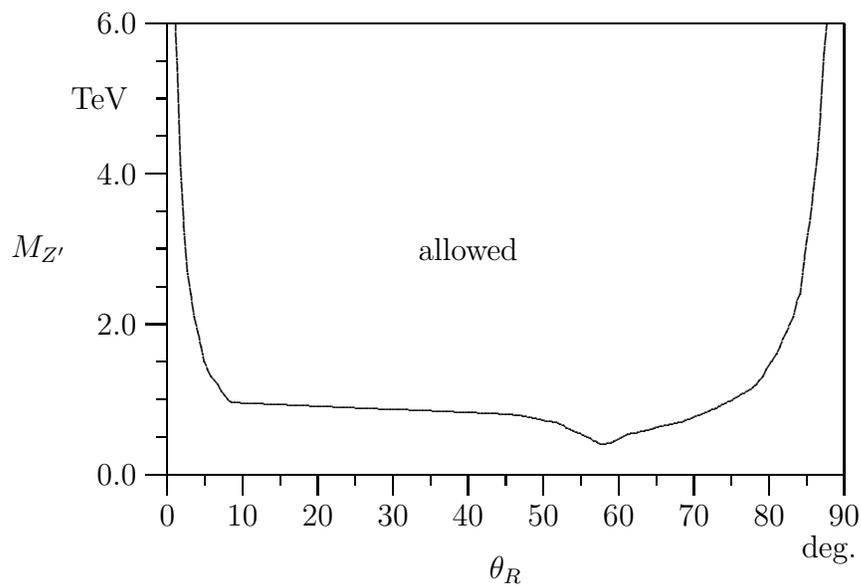

\vspace{0.4in}
\beginpicture
\setcoordinatesystem units <0.1cm,0.1cm> 
\setplotarea x from 0 to 90, y from 0 to 60
\axis bottom label {$\theta_R$} ticks
length <4pt> unlabeled from 0 to 90 by 5.0
length <8pt> width <.4pt> withvalues  0 10 20 30 40 50
60 70 80  90 / 
at 0.0 10.0 20.0 30.0 40.0 50.0 60.0 70.0 80.0  90.0  / /
\put {deg.} at 88 -10
\axis top  /
\axis left label {$M_{Z^{\prime}}$} ticks
length <4pt> unlabeled from 0 to 60 by 5.0 
length <8pt> width <.4pt> withvalues 0.0 2.0 4.0 6.0  /
at 0.0 20.0 40.0 60.0 / /
\axis right ticks /
\put {TeV} at -9 50
\put {allowed} at 40 30
\setlinear
\plot 
 1.1    60.0
 1.38   54.0
 1.78   41.0
 2.23   33.0
 2.69   27.0
 3.15   24.0
 3.61   21.0
 4.07   19.0
 4.53   17.0
 4.93   15.0
 5.39   14.0
 5.84   13.0
 6.30   12.5
 6.76   12.0
 7.22   11.0
 7.68   10.5
 8.08   10.00
 8.54   9.6
 9.00   9.54
 40.05   8.24
 40.51   8.22
 40.97   8.2
 41.43   8.18
 41.83   8.16
 42.29   8.14
 42.75   8.12
 43.20   8.1
 43.66   8.08
 44.12   8.06
 44.58   8.04
 44.98   8.02
 45.44   8.0
 45.90  7.95
 46.36  7.91
 46.81   7.9
 47.27   7.8
 47.73   7.7
 48.13   7.6
 48.59   7.5
 49.05   7.4
 49.51   7.3
 49.97   7.2
 50.43   7.1
 50.83   7.05
 51.29   7.0
 51.75  6.9
 52.20   6.7
 52.66   6.5
 53.12   6.2
 53.58   6.0
 53.98   5.8
 54.44   5.6
 54.90   5.4
 55.36   5.2
 55.81   5.0
 56.27   4.8
 56.73   4.5
 57.13   4.3
 57.59   4.0
 58.05   4.0
 58.51   4.1
 58.97   4.2
 59.43   4.4
 61.20   5.3
 61.66   5.4
 62.12   5.5
 62.58   5.6
 62.98   5.7
 63.44   5.8
 63.90   5.9
 64.36   6.0
 64.81   6.2
 65.27   6.3
 65.73   6.4
 66.13   6.5
 66.59   6.6
 67.05   6.7
 67.51   6.8
 67.97   6.9
 68.43   7.0
 68.87   7.2
 69.33   7.3
 69.73   7.5
 70.19   7.7
 70.65   7.9
 71.11   8.0
 71.57   8.2
 72.03   8.4
 72.43   8.6
 72.89   8.8
 73.34   9.0
 73.80   9.3
 74.26   9.5
 74.72   9.7
 77.87   11.5
 78.33   12.0
 78.73   12.5
 79.19   13.0
 79.65   14.0
 81.02   16.0
 81.48   17.0
 81.88   18.0
 82.34   19.0
 82.80   20.00
 83.26   21.0
 83.72   23.0
 84.17   24.0
 84.63   28.0
 85.03   31.0
 85.49   34.0
 85.95   38.0
 86.41   42.0
 86.87   48.0
 87.33   56.0
 87.7   60.0 /
\endpicture
\caption{Bounds on $\theta_R$ and $M_{Z^{\prime}}$ satisfying the
low-energy neutral-current data and the LEP I constraints at 95\%
CL. The region above the curve is allowed.} 
\label{fig3} 
\end{figure}

\begin{figure}[ht]
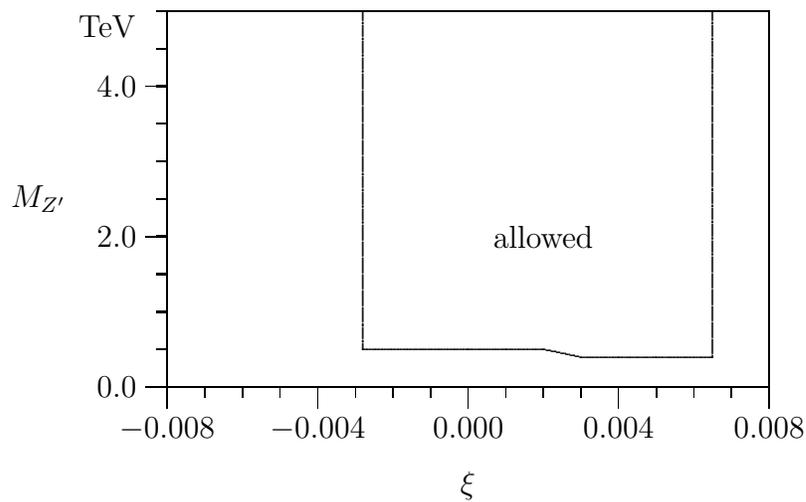

\vspace{0.4in}
\beginpicture
\setcoordinatesystem units <0.1cm,0.1cm> 
\setplotarea x from -40 to 40, y from 0 to 50
\axis bottom label {$\xi$} ticks
length <8pt> width <.2pt> unlabeled from -40 to 40 by 20.0
length <4pt> unlabeled from -40 to 40 by 5.0
withvalues $-$0.008 $-$0.004 0.000  0.004 0.008  /
at -40 -20 0 20 40  / /
\axis top /
\axis left label {$M_{Z^{\prime}}$} ticks
length <4pt> unlabeled from 0 to 50 by 5.0 
length <8pt> width <.2pt> withvalues 0.0 2.0 4.0  /
at 0.0 20.0 40.0  / /
\axis right /
\put {TeV} at -48 48
\put {allowed} at 10 20
\setlinear
\plot  -14.0  50.0  -14.0  5.0
  0.  5.0  5.0  5.0  10.0 5.0 15.0 4.0 20.0 4.0 25.0 4.0 32.5 4.0 
32.5 50.0 /
\endpicture
\caption{Bounds on $\xi$ and $M_{Z^{\prime}}$ satisfying the
low-energy neutral-current data and the LEP I constraints at 95\% 
CL.} 
\label{fig4}
\end{figure}

We vary $\theta_R$ freely since there is little constraint from the
LEP I data. However, since the bounds for $\xi$ depend on $\theta_R$,
we vary $\xi$ according to the relation with $\theta_R$ as shown in
Fig.~\ref{fig1}. We also vary $M_{Z^{\prime}}$ and look for the values
of the set $(\theta_R, \xi, M_{Z^{\prime}})$ which satisfy the
low-energy neutral-current data. In order to show the bounds in a
transparent way, we consider the bounds in two-parameter spaces
$(\theta_R, M_{Z^{\prime}})$ and $(\xi, M_{Z^{\prime}})$. We do not
show the bounds in the $(\xi,\theta_R)$ space since the low-energy
data do not put additional constraints on the bounds from LEP I.

First, we show the bounds in the $(\theta_R, M_{Z^{\prime}})$ space 
in Fig.~\ref{fig3}. As can be seen in Fig.~\ref{fig3}, the low-energy
data do not constrain $\theta_R$ either. However, we obtain the bound
for $M_{Z^{\prime}}$,
\begin{equation}
M_{Z^{\prime}} > 400 \ \mbox{GeV}.
\end{equation}
The smallest lower bound occurs at $\theta_R = 55^{\circ} -
60^{\circ}$ and the bound increases when $\theta_R$ is away from this 
region. The left-right symmetric case corresponds to $\theta_R =
33^{\circ}$ and the lower bound at the point is about 900 GeV. In
another parameter space spanned by $(\xi, M_{Z^{\prime}})$, the result
is shown in Fig.~\ref{fig4}.  

In summary, the analysis using the combined data from LEP I and the
low-energy experiments shows that the bounds are given as 
\begin{equation}
-0.0028 < \xi < 0.0065, \ \ M_{Z^{\prime}} > 400 \ \mbox{GeV},
\end{equation}
with little constraint on $\theta_R$. 

Before finishing this analysis, there is one technical comment about
dealing with the accuracy of experimental data. We fit the parameters
to the experimental data at 95\% CL. If we analyze the data at 90\%
CL, there are more stringent limits on the parameters. The LEP
I data include the standard model value at 90\% CL, so there is no
problem in fitting the parameters at 90\% CL for LEP I data. However,
for low-energy neutral-current data, the standard model value for
$\epsilon_R (u)$ only is outside the 90\%-CL estimates. Therefore at
90\% CL, the most severe constraint comes from $\epsilon_R (u)$ and in
this case, the bounds are given by 
\begin{equation}
\theta_R > 72^{\circ}, \\ -0.0014 <\xi < 0.0003, \ \ 800 \ \mbox{GeV}
< M_{Z^{\prime}} < 8.3 \ \mbox{TeV}.
\end{equation}
In this case, the left-right symmetric case is disfavored. However,
we need more data to obtain precise central values of the experimental
data and to decrease experimental errors in order to draw a
definite conclusion. 

\subsection{Comparison with LEP II data}
We consider the total cross section and the forward-backward
asymmetry for $e^+e^- \rightarrow \mu^+ \mu^-$ at $s > M_Z^2$. The
differential cross section is written as 
\begin{equation}
\frac{d\sigma}{d\cos\theta} = \frac{1}{32\pi s} \Bigl( S(1+\cos^2
\theta) + 2A \cos \theta\Bigr),
\end{equation}
where $\theta$ is the angle between the incoming electron and the
outgoing muon in the center-of-mass frame. The total cross section is
given by \begin{equation}
\sigma = \sigma_F + \sigma_B = \frac{1}{12\pi s} S(1+\Delta_{\ell}), 
\end{equation}
where $\Delta_{\ell}$ is the one-loop electroweak correction.
And the forward-backward asymmetry can be expressed as 
\begin{equation}
A_{FB} = \frac{\sigma_F -\sigma_B}{\sigma_F+\sigma_B} = \frac{3}{4}
\frac{A}{S}. 
\end{equation}

We can express $\sigma$ and $A_{FB}$ to first order in $\xi$ and
$M_Z^2/M_{Z^{\prime}}^2$. The coupling of the charged leptons with $Z$
and $Z^{\prime}$ are obtained to first order in $\xi$ in the limit of
small mixing angle $\xi$. If we write the leptonic current coupled to 
$Z$ as  $-(e/2s_W c_W) \overline{\ell} \gamma^{\mu}
(g_{\mathrm{VZ}} -g_{\mathrm{AZ}} \gamma_5) \ell$, $g_{\mathrm{VZ}}$
and $g_{\mathrm{AZ}}$ are given by  
\begin{equation}
g_{\mathrm{VZ}} =  g_V + \xi s_W ( t_R - \frac{1}{2t_R}), \ \ 
g_{\mathrm{AZ}} = g_A +\xi \frac{s_W}{2t_R}.
\end{equation}
Here $g_V = -\half + 2s_W^2$ and $g_A = -\half$ are the couplings in
the standard model. Similarly, if we write the
coupling with $Z^{\prime}$  as $-(e/2s_W c_W) \overline{\ell}
\gamma^{\mu} (g_{\mathrm{VZ^{\prime}}} - g_{\mathrm{AZ^{\prime}}}
\gamma_5) \ell$, the coupling constants are given by
\begin{equation} 
g_{\mathrm{VZ^{\prime}}} = s_W (t_R - \frac{1}{2t_R}), \ \ 
g_{\mathrm{AZ^{\prime}}} = \frac{s_W}{2t_R}.
\end{equation}

For $e^+ e^- \rightarrow \mu^+ \mu^-$, $A$ and $S$, to first
order in the mixing angle $\xi$ and $M_Z^2/M_{Z^{\prime}}^2$, are
given as  
\begin{eqnarray} 
A &=& \frac{e^4}{4s_W^4 c_W^4} 
\frac{s^2}{(s-M_Z^2)^2} g_V^2 g_A^2+
\frac{e^4}{2s_W^2 c_W^2} \frac{s}{s-M_Z^2} g_A^2\nonumber
\\ 
&+& \xi \Bigl[ \frac{e^4}{2s_W^3 c_W^4} 
\frac{s^2}{(s-M_Z^2)^2} g_Vg_A \Bigl( g_A t_R +  \frac{g_V -
g_A}{2t_R} \Bigr) +\frac{e^4}{2s_W c_W^2} \frac{s}{s-M_Z^2}
\frac{g_A}{t_R} \Bigr] \nonumber \\
&-& \frac{s}{M_{Z^{\prime}}^2} \Bigl[ \frac{e^4}{8 c_W^2}
\frac{1}{t_R^2}  +\frac{e^4}{8s_W^2 c_W^4} \frac{s}{s-M_Z^2} 
\Bigl( g_A t_R + \frac{g_V -
g_A}{2t_R} \Bigr) \Bigr], 
\end{eqnarray}
and
\begin{eqnarray}
S &=& e^4 + \frac{e^4}{16s_W^4 c_W^4}
\frac{s^2}{(s-M_Z^2)^2} (g_V^2 + g_A^2)^2+
\frac{e^4}{2s_W^2 c_W^2} 
\frac{s}{s-M_Z^2} g_V^2 \nonumber \\
&+& \xi \Bigl[ \frac{e^4}{4s_W^3 c_W^4} 
\frac{s^2}{(s-M_Z^2)^2} (g_V^2 +g_A^2)\Bigl( g_V t_R +
\frac{g_A - g_V}{2t_R} \Bigr) \nonumber \\
&+& \frac{e^4}{s_W c_W^2} \frac{s}{s-M_Z^2} g_V (t_R
-\frac{1}{2t_R}) \Bigr] \nonumber \\
&-& \frac{s}{M_{Z^{\prime}}^2} \Bigl[ \frac{e^4}{2 c_W^2} (t_R -
\frac{1}{2t_R})^2 
+\frac{e^4}{8s_W^2 c_W^4} \frac{s}{s-M_Z^2} \Bigl
( g_V t_R +\frac{g_A -
g_V}{2t_R} \Bigr)^2 \Bigr].
\end{eqnarray}

For $e^+ e^- \rightarrow b\overline{b}$, we can similarly parameterize
the cross section. The quark current coupled with $Z$ is given by
$-(e/2s_W c_W) \overline{b} \gamma^{\mu} ( g_{\mathrm{VZ}}^b -
g_{\mathrm{AZ}}^b \gamma_5)b$ where 
\begin{equation}
g^b_{\mathrm{VZ}}= g^b_V
-\xi s_W (\frac{t_R}{3} +\frac{1}{2t_R}), \ \ 
g^b_{\mathrm{AZ}}= g^b_A + \xi s_W \frac{1}{2t_R},
\end{equation}
and $g^b_V = -\frac{1}{2} + \frac{2}{3}s_W^2$, and $g^b_A =
-\frac{1}{2}$ are the standard model values. 

The total cross section is given by  
\begin{equation}
\sigma_b = \frac{1}{12\pi s} S_b (1+\Delta_b),
\end{equation}
where
\begin{eqnarray}
S_b &=& \frac{1}{9} \frac{e^4}{16 s_W^4 c_W^4}
\frac{s^2}{(s-M_Z^2)^2} (g_V^{b2} + g_A^{b2})^{2} 
+ \frac{e^4}{6s_W^2 c_W^2} \frac{s}{s-M_Z^2} g_V^{b2} 
\nonumber \\
&+& \xi \Bigl[ \frac{e^4}{4s_W^3 c_W^4} \frac{s^2}{(s-M_Z^2)^2}
(g_V^{b2} + g_A^{b2}) \Bigl ( -\frac{g^b_V}{3}  
+\frac{g^b_A -g^b_V}{2t_R} \Bigr) \nonumber \\ 
&-& \frac{e^4}{3s_W c_W^2} \frac{s}{s-M_Z^2} g^b_V 
\Bigl( \frac{t_R}{3} + \frac{1}{2t_R} \Bigr) \Bigr]\nonumber \\
&-& \frac{s}{M_{Z^{\prime}}^2} \Bigl[ \frac{e^4}{6c_W^2} \Bigl
( \frac{t_R}{3} + \frac{1}{2t_R} \Bigr)^2 
+ \frac{e^4}{8s_W^2 c_W^4} \frac{s}{s-M_Z^2} \Bigl
( -\frac{g^b_V}{3} t_R + \frac{g^b_A -
g^b_V}{2t_R} \Bigr)^2 \Bigr].
\end{eqnarray}
And the QCD correction factor in $\Delta_b$ is given by 
\begin{equation}
\Delta_{\mathrm{QCD}} = 1.2 \frac{\alpha_s (\sqrt{s})}{\pi} - 1.1
\Bigl( \frac{\alpha_s (\sqrt{s})}{\pi} \Bigr)^2 + \cdots.
\label{bdelta}
\end{equation}

\begin{table}
\begin{tabular}{cccccc} \hline
& $\sqrt{s}$ & & 130.12 GeV & 136.08 GeV & 183 GeV \\  \hline
& $A_{FB}^l$ & SM & 0.70 & 0.68 & 0.57 \\
& & LR model & 0.60 -- 0.81 & 0.57 -- 0.81 & 0.34 -- 0.79 \\
& & Experiments & 0.55 $\pm$ 0.13 & 0.76$ \pm$ 0.09 & 0.60 $\pm$
0.05\\ \hline
& $\sigma^\mu$ & SM & 8.5 & 7.3 & 3.45 \\
& & LR model & 6.93 -- 8.52 & 5.78 -- 7.32 & 2.07 -- 3.46 \\
& & Experiments & 7.6 $\pm$ 1.4 & 10.4 $\pm$ 1.6 & 3.46 $\pm$ 0.38\\
\hline
& $\sigma^b$ & SM &  &  & 3.96 \\
& & LR model &  &  & 3.59 -- 3.97 \\
& & Experiments & -- &--  & 4.6 $\pm$ 0.9 \\ \hline
\end{tabular}
\vspace{0.2in}
\caption{Comparison of the LEP II data, the standard model (SM), and
the $SU(2)_L\times SU(2)_R\times U(1)$ (LR) model. The cross
sections and the forward-backward asymmetries at different energies
are listed in each case.} 
\label{table2}
\vspace{0.3in}
\end{table}

We compare our results with those of the OPAL Collaboration
\cite{opal}. The accuracy of the experimental data from LEP II is not
as good as that from LEP I, but it will be improved as more data will
be accumulated. We find that the bounds using the LEP II data do not
give more severe bounds obtained from LEP I. Instead, we calculate the
cross sections and the forward-backward asymmetry at LEP II using the
bounds obtained from the combined data of LEP I and the low-energy
data.

The results are shown in Table~\ref{table2}. The cross sections
and the forward-backward asymmetry at different energies are
listed. The results from the $SU(2)_L \times SU(2)_R \times U(1)$ (LR)
model and the standard model (SM) are shown along with the current
experimental data. The standard model values, quoted in Table~2,
include the effects of the electroweak radiative corrections and the
QCD corrections. Most of the values in our model are within 1$\sigma$
of the  experimental values, but the comparison will be useful after
the experimental results are more refined.

\section{Conclusion}
We have studied constraints on the neutral sector in the $SU(2)_L
\times SU(2)_R \times U(1)$ model. We introduce three mixing angles
$\xi$, $\theta_R$ and $\theta_W$ to diagonalize the neutral gauge
boson mass matrix. Here $\theta_W$ is identified as the Weinberg
angle and we use the remaining two mixing angles and the heavy neutral
gauge boson mass $M_{Z^{\prime}}$ to describe new physics effects from
the neutral sector in the model. Since $\xi$ and
$M_Z^2/M_{Z^{\prime}}^2$ are small parameters, we calculate all the
corrections to the standard model in various processes to first order
in these small parameters and fit to experimental data. 

First we use the LEP I data to constrain $\xi$ and  $\theta_R$ without
any information on $M_{Z^{\prime}}$ since the LEP I energy is at the
$Z$ peak. There is little constraint on $\theta_R$, but $\xi$ is
bounded by $-0.0028 < \xi < 0.0065$ for all $\theta_R$. Note that the
bound for $\xi$ varies for different values of $\theta_R$ as shown in 
Fig.~\ref{fig1}.  With the constraints obtained from the LEP I data,
we find the bounds for $\xi$ and $M_{Z^{\prime}}$ which simultaneously
satisfy the low-energy neutral-current data. The combined bounds at
95\% CL are given as 
\begin{equation}
-0.0028 < \xi < 0.0065, \ \ M_{Z^{\prime}} > 400 \ \mbox{GeV}.
\end{equation}
The bound for the mixing angle $\xi$ in the neutral sector is more
severe compared to the bound for  the mixing angle $\zeta$ in the
charged sector, $|\zeta| < 0.075$. We also consider other experimental
results such as the LEP II data in the context of the $SU(2)_L \times
SU(2)_R \times U(1)$ model.

The lower bound for the $Z^{\prime}$ mass is 400 GeV when we combine
the LEP I data and the low-energy neutral-current data. In the case of
the left-right symmetric model with $g_L = g_R$ ($\theta_R =
33^{\circ}$), $M_{Z^{\prime}}> 900$ GeV. On the other hand, we have
considered the relation between $M_{W^{\prime}}$ and $M_{Z^{\prime}}$
in Sec. 3.1. In the left-right symmetric case, it is given  as 
\begin{equation}
M_{Z^{\prime}} =  \frac{M_{W^{\prime}}}{\cos \theta_R} >
\frac{M_{W^{\prime}}}{\cos \theta_W},
\end{equation}
using the fact that $\cos \theta_R < \cos \theta_W$
for $\theta_R = 33^{\circ}$. If we accept the assumptions and the
result in Ref.~\cite{ecker} for the left-right symmetric theory, 
we can get a more severe bound   
\begin{equation}
M_{Z^{\prime}} > 1.6 \ \mbox{TeV}
\label{lrmz}
\end{equation}
for $M_{W^{\prime}} > 1.4$ TeV. Note that this result is obtained from
an independent information on $M_{W^{\prime}}$, while the bound from
the analysis of the neutral sector is $M_{Z^{\prime}}> 900$
GeV. However, when we consider the bound in Eq.~(\ref{lrmz}), there is
a caveat that the bound $M_{W^{\prime}}$ from the charged sector
depends on many assumptions. 

There has been a search for an additional $Z^{\prime}$ particle
irrespective of the detailed structure of the theory. From the search
for the process $Z^{\prime} \rightarrow \mu^+ \mu^-$ at Fermilab
\cite{fermilab}, the bound for $M_{Z^{\prime}}$ at 95\% CL is
$M_{Z^{\prime}} > 412$ GeV. This is the result independent of the
experimental data considered here. However, it is interesting to note
that the lower bounds for $M_{Z^{\prime}}$ in both cases are similar. 

Amaldi et al. and Costa et al.\cite{neutral} considered the model
with an additional $U(1)$ from string-inspired models. They obtained
the limits on the  $Z^{\prime}$ mass larger than 325 GeV, and the
mixing angle corresponding $\xi$ in our model should satisfy $|\xi| <
0.05$ considering low-energy data. Cho et al. \cite{cho} have
considered the additional neutral particle $Z^{\prime}$ in the context
of the supersymmetric $E_6$ models. Their results are based on
the heavy $Z^{\prime}$ from an additional $U(1)$ gauge
group. Therefore care should be taken in comparing their results with
ours. 

It is also interesting to get bounds on these parameters from other
experiments such as $B$ decays. Cho and Misiak \cite{misiak} have
considered the decay rate for $b\rightarrow s\gamma$ in the $SU(2)_L
\times SU(2)_R \times U(1)$ models. Their conclusion is that though
QCD corrections diminish the difference between this model and the
standard model, but for reasonable ranges of parameters, the decay
rates can be distinguished and used to probe for new physics beyond
the standard model. Babu et al. \cite{babu} have considered the same
process including the effect of the Higgs particle exchange and
obtained the result $-0.015 < \zeta < 0.003$ and $M_H>$ a few GeV. 
However, in the decay $b\rightarrow s\gamma$, only
the charged gauge bosons contribute to the process. In order to
consider the new physics effects from the neutral sector, it may be
interesting to consider decays such as $b\rightarrow sl^+ l^-$.
The search for bounds on the parameters in the $SU(2)_L \times SU(2)_R 
\times U(1)$ model in $B$ decays such as $B\rightarrow X_s \ell^+
\ell^-$ or $B\rightarrow X_s \overline{\nu} \nu$ is in progress
\cite{chaylee}.  

\begin{ack}
All the authors are supported by the Korea Science and Engineering
Foundation (KOSEF) through the SRC program of SNU-CTP. JC is supported
in part by the Ministry of Education grants BSRI 98-2408,
the Distinguished Scholar Exchange Program of
Korea Research Foundation and the German-Korean scientific exchange
program DFG-446-KOR-113/72/0.
\end{ack}

\section*{Appendix: Diagonalization of the Neutral Gauge Boson Masses}
Here we show in detail how to diagonalize the neutral
gauge boson mass matrix. In order to obtain physical gauge boson
states, we have to diagonalize the mass matrix in
Eq.~(\ref{neumass}). Since the mass-squared matrix is a real,
symmetric $3\times 3$ matrix, we need a real, orthogonal matrix to
diagonalize it. That is, we have to find three Euler angles
parameterizing the orthogonal matrix which diagonalizes the mass
matrix. There are two ingredients to facilitate the
diagonalization. First we assume that the VEV $v_R$ is much larger 
than other VEVs such as $v_L$, $|k|$, $|k^{\prime}|$. And since the
electromagnetic $U(1)_{\mathrm{em}}$ remains unbroken, there
remains one massless field which corresponds to the photon field.

As a first step, consider the mass matrix in Eq.~(\ref{neumass}) in
the limit $v_R \rightarrow \infty$ and neglect small terms compared to
$v_R$.  Then the mass-squared matrix is written as 
\begin{equation}
M^2 \approx \left( \begin{array}{ccc}
0&0&0 \\
0&g_R^2 v_R^2 /2& -g_R g_1 v_R^2 /2 \\
0& -g_R g_1 v_R^2/2& g_1^2 v_R^2 /2
	     \end{array}
\right).
\end{equation}
We can diagonalize the lower right $2\times 2$ block matrix by
defining the following fields:
\begin{equation}
Z_1 = B\cos \theta_R + W_{R3}\sin \theta_R , \ 
\hat{Z} = -B\sin \theta_R + W_{R3} \cos \theta_R,
\label{zone}
\end{equation}
where
\begin{equation}
\frac{g_R}{\sqrt{g_1^2 + g_R^2}} = \cos \theta_R, \
\frac{g_1}{\sqrt{g_1^2 + g_R^2}} = \sin  \theta_R.
\end{equation}
If we rewrite the mass matrix in the basis of $W_{L3}$, $Z_1$ and
$\hat{Z}$, the biggest VEV, $v_R$, appears only in the lower right
end. 

Now we use the fact that one of the eigenvalues of the matrix in
Eq.~(\ref{neumass}) is zero, which corresponds to the photon field. It
is straightforward to obtain the eigenvector with the eigenvalue 0. It
is given by  
\begin{equation}
A= \frac{g_1 g_R W_{L3} + g_L \sqrt{g_1^2 + g_R^2} Z_1}{\sqrt{g_1^2
g_R^2 + g_L^2 g_1^2 + g_L^2 g_R^2}}.
\end{equation}
Let us define another field $\tilde{Z}$ which is orthogonal to the
photon field $A$:
\begin{equation}
A= \sin \theta_W W_{L3} +\cos \theta_W Z_1, \ \ 
\tilde{Z} = \cos \theta_W W_{L3} -\sin \theta_W Z_1,
\label{aandz}
\end{equation}
where 
\begin{equation}
\frac{g_1 g_R}{\sqrt{g_1^2 g_R^2 + g_L^2 g_1^2 + g_L^2 g_R^2}} = \sin
\theta_W, \ \frac{g_L \sqrt{g_1^2 + g_R^2}}{\sqrt{g_1^2 g_R^2 + g_L^2
g_1^2 + g_L^2 g_R^2}} = \cos \theta_W.
\end{equation}
The mixing angle $\theta_W$ corresponds to the Weinberg mixing angle
in the standard model if we identify $Z_1$ as the neutral $Z$ in the
standard model. Actually $\theta_W$ is equal to the Weinberg
mixing angle in the limit $v_R \rightarrow \infty$ and the correction
is of order $O(v_L^2/v_R^2)$.

In the basis of $A$, $\tilde{Z}$ and $\hat{Z}$, the mass-squared matrix
takes a simple form. Nonzero terms appear only in the lower right
$2\times 2$ block. The mass-squared matrix looks like 
\begin{equation}
M^2 = \left( \begin{array}{ccc}
0&0&0\\
0&M_{\tilde{Z}}^2 & M_{\tilde{Z} \hat{Z}}^2 \\
0&M_{\tilde{Z} \hat{Z}}^2& M_{\hat{Z}}^2
	     \end{array}
\right),
\end{equation}
where
\begin{eqnarray}
M_{\tilde{Z}}^2 &=& \frac{1}{2} \frac{g_L^2}{c_W^2} (K^2 +
v_L^2), \ \
M_{\hat{Z}}^2 = \frac{1}{2} (g_1^2 + g_R^2) \bigl( v_R^2 + c_R^4 K^2
+  s_R^4 v_L^2 \bigr), \nonumber \\ 
M_{\tilde{Z} \hat{Z}}^2&=& -\frac{1}{2} \frac{g_L}{c_W} 
g_R c_R (K^2 - t_R^2 v_L^2),
\label{nmass}
\end{eqnarray}
where $t_R = s_R/c_R$. Note that $v_R$ appears only in
$M_{\hat{Z}}^2$. Therefore we expect that the mixing angle to
diagonalize this matrix is small.

Now the diagonalization of the remaining $2\times 2$ can be done in a
similar way to diagonalize the charged gauge boson masses. We
introduce the mixing angle as
\begin{equation}
\tan 2\xi = -\frac{2M_{\tilde{Z} \hat{Z}}^2}{M_{\hat{Z}}^2 -
M_{\tilde{Z}}^2}. 
\label{xirel}
\end{equation}

Finally the physical neutral gauge bosons $Z$ and $Z^{\prime}$ can be
written as
\begin{equation}
Z = \tilde{Z} \cos \xi + \hat{Z} \sin \xi, \ Z^{\prime} = -\tilde{Z}
\sin \xi + \hat{Z} \cos \xi. 
\end{equation}
The field $Z$ corresponds to the $Z$ gauge boson in the standard model
and the field $Z^{\prime}$ is a new field which is more massive than
the $Z$ particle. The corresponding mass eigenvalues are 
\begin{eqnarray} 
M_Z^2 &=& M_{\tilde{Z}}^2 \cos^2 \xi +M_{\hat{Z}}^2 \sin^2 \xi
+M_{\tilde{Z} \hat{Z}}^2 \sin 2\xi, \nonumber \\
M_{Z^{\prime}}^2 &=& M_{\tilde{Z}}^2 \sin^2 \xi + M_{\hat{Z}}^2 \cos^2 \xi
-M_{\tilde{Z} \hat{Z}}^2 \sin 2\xi.
\label{zmass}
\end{eqnarray}
In the limit $v_R \gg v_L, |k|, |k^{\prime}|$, these are approximately
given by 
\begin{eqnarray}
M_Z^2 &\approx& \frac{1}{2} \frac{g_L^2}{c_W^2} (K^2 + v_L^2)
-\frac{g_L^2}{2c_W^2} \frac{(c_R^2 K^2 - s_R^2 v_L^2)}{v_R^2},
\nonumber \\  
M_{Z^{\prime}}^2 &\approx& \frac{1}{2} (g_1^2 + g_R^2) v_R^2 +
\frac{1}{2} (g_R^2 c_R^2 K^2 +g_1^2 s_R^2 v_L^2).
\end{eqnarray}
Note that with a fine tuning $K = t_R v_L$, the mixing term
$M_{\tilde{Z} \hat{Z}}^2$ becomes zero, hence $\xi =0$. This is one of
the cases we consider in constraining the remaining parameters.

In summary, the physical mass eigenstates $A$, $Z$ and $Z^{\prime}$
fields can be written as a linear combination of the gauge eigenstates
$W_{L3}$, $W_{R3}$ and $B$. Since the $3\times 3$ mass-squared matrix
is a real, symmetric matrix, we need an orthogonal matrix to
diagonalize it. In the prescription described above, we find three
mixing angles which are Euler angles to parameterize the orthogonal
matrix. The physical fields can be written as
\begin{eqnarray}
A &=& s_W W_{L3} + c_W s_R W_{R3} + c_W c_R B, \nonumber \\
Z&=& c_W c_{\xi} W_{L3} +(c_R s_{\xi} -s_W s_R c_{\xi}) W_{R3}
-(s_W c_R c_{\xi} + s_R s_{\xi}) B, \nonumber \\
Z^{\prime} &=& -c_W s_{\xi} W_{L3} +(c_R c_{\xi} + s_W s_R s_{\xi})
W_{R3} +(s_W c_R s_{\xi} -s_R c_{\xi})B,
\label{rot}
\end{eqnarray}
where $c_{\xi} = \cos \xi$ and $s_{\xi} = \sin \xi$.

It is useful to verify the result by taking the limit $g_L = g_R$. In
this case $\sin \theta_R = \tan \theta_W$ and $\cos \theta_R =
\sqrt{\cos 2\theta_W}/\cos \theta_W$. Using these relations, and
taking the limit $\xi \rightarrow 0$, Eq.~(\ref{rot}) becomes
\begin{eqnarray}
A &=& s_W (W_{L3} + W_{R3}) + \sqrt{\cos 2\theta_W} B, \nonumber \\ 
Z&\approx& c_W  W_{L3} -s_W t_W  W_{R3}
-t_W \sqrt{\cos 2\theta_W}  B, \nonumber \\
Z^{\prime} &\approx& \frac{\sqrt{\cos 2\theta_W}}{c_W} W_{R3} -t_W B.
\label{rotlim}
\end{eqnarray}
This coincides with the symmetric result of Ref.~\cite{senja}

\end{document}